\documentclass{article}

\usepackage{arxiv}

\usepackage[utf8]{inputenc} 
\usepackage[T1]{fontenc}    
\usepackage{hyperref}       
\usepackage{url}            
\usepackage{booktabs}       
\usepackage{amsfonts}       
\usepackage{nicefrac}       
\usepackage{microtype}      
\usepackage{lipsum}
\usepackage{graphicx}
\graphicspath{ {./images/} }

\usepackage{amsthm}
\usepackage{amsmath}
\usepackage{xfrac}
\usepackage{amssymb}
\usepackage{algorithm}
\usepackage{algpseudocode}
\usepackage{bm}

\usepackage[numbers]{natbib} 

\title{$\aleph$-IPOMDP: Mitigating Deception in a Cognitive Hierarchy with Off-Policy Counterfactual Anomaly Detection\thanks{Accepted for publication in the Journal of Artificial Intelligence Research (JAIR), January 2026.}}

\renewcommand{\shorttitle}{$\aleph$-IPOMDP: Mitigating Deception}

\author{
  Nitay Alon\thanks{Corresponding author: nitay.alon@tuebingen.mpg.de} \\
  Department of Computational Neuroscience\\
  Max Planck Institute for Biological Cybernetics\\
  T\"{u}bingen, Germany \\
  \And
  Joseph M. Barnby \\
  Centre for AI and Machine Learning\\
  Edith Cowan University\\
  Perth, Australia \\
  \And
  Stefan Sarkadi \\
  Centre for Defence and Security Artificial Intelligence \\
  University of Lincoln\\
  Lincoln, UK \\
  \And
  Lion Schulz \\
  Department of Computational Neuroscience\\
  Max Planck Institute for Biological Cybernetics\\
  T\"{u}bingen, Germany \\
  \And
  Jeffrey S. Rosenschein \\
  School of Computer Science and Engineering\\
  The Hebrew University of Jerusalem\\
  Jerusalem, Israel \\
  \And
  Peter Dayan \\
  Department of Computational Neuroscience\\
  Max Planck Institute for Biological Cybernetics\\
  T\"{u}bingen, Germany \\
}

\begin{document}
\maketitle

\begin{abstract}
Social agents with finitely nested opponent models are vulnerable to manipulation by agents with deeper recursive capabilities. This imbalance, rooted in logic and the theory of recursive modelling frameworks, cannot be solved directly. We propose a computational framework called $\aleph$-IPOMDP, which augments the Bayesian inference of model-based RL agents with an anomaly detection algorithm and an out-of-belief policy. Our mechanism allows agents to realize that they are being deceived, even if they cannot understand how, and to deter opponents via a credible threat. We test this framework in both a mixed-motive and a zero-sum game. Our results demonstrate the $\aleph$-mechanism's effectiveness, leading to more equitable outcomes and less exploitation by more sophisticated agents. We discuss implications for AI safety, cybersecurity, cognitive science, and psychiatry.
\end{abstract}


\section{Introduction}
Deception is constant in human and animal cultures. Humans use a range of deceptive techniques, from ``white lies'' to malicious and harmful manipulation, misdirecting the beliefs of others for their benefit. To manipulate, a deceiver needs to both create false beliefs and avoid disclosing their true intentions. Agents can achieve this through perspective taking, a form of Theory of Mind \citep[ToM;][]{premack_does_1978}. ToM encompasses the capacity to simulate others' actions and beliefs. This can be shallow, e.g., observational learning, or recursive (i.e., including the other's capacity to simulate the self, and so forth). The degree to which an agent can use recursive beliefs is known as its depth of mentalising \citep[DoM;][]{barnby2023formalising,frith2021mapping}. This property has been explained through formal models. A successful and popular framing is k-level hierarchical ToM \citep{camerer_cognitive_2004}. K-level ToM predicts that agents with lower DoM are formally incapable of making accurate inferences about the intentions of those with higher DoM \citep{gmytrasiewicz_framework_2005}. Such an ability would suggest that agents had circumvented the paradox of self-reference. This limitation, found in all recursive modelling frameworks \citep{pacuit_epistemic_2017}, implies that agents with low DoM are doomed to be manipulated by others with higher DoM. This asymmetry has previously been explored \citep{de_weerd_higher-order_2022, hula_model_2018, alon2023between, sarkadi_evolution_2021, sarkadi_modelling_2019}, illustrating the various ways that higher DoM agents can take advantage of lower DoM interaction partners. While these agents are necessarily disadvantaged, all is not lost. Low DoM agents may still notice that the behaviour they \emph{observe} is inconsistent with the behaviour they \emph{expect}, even if they lack the knowledge to understand how or why \citep{hula_model_2018}. This type of mismatch warns the victim that they are facing an unmodeled opponent, meaning they can no longer use their simulation of an other for optimal planning. One way to combat this apparent model failure is to switch to an out-of-belief (OOB) policy, where actions are detached from inferences about their opponent. One path from this is to switch from exploit to exploration behaviour, opening the way to the development of a new opponent model. However, if losing carries a detrimental impact, another option may be to take defensive action, choosing to avoid or reject an environment.

\begin{figure}[ht]
    \centering
    \includegraphics[scale=.45]{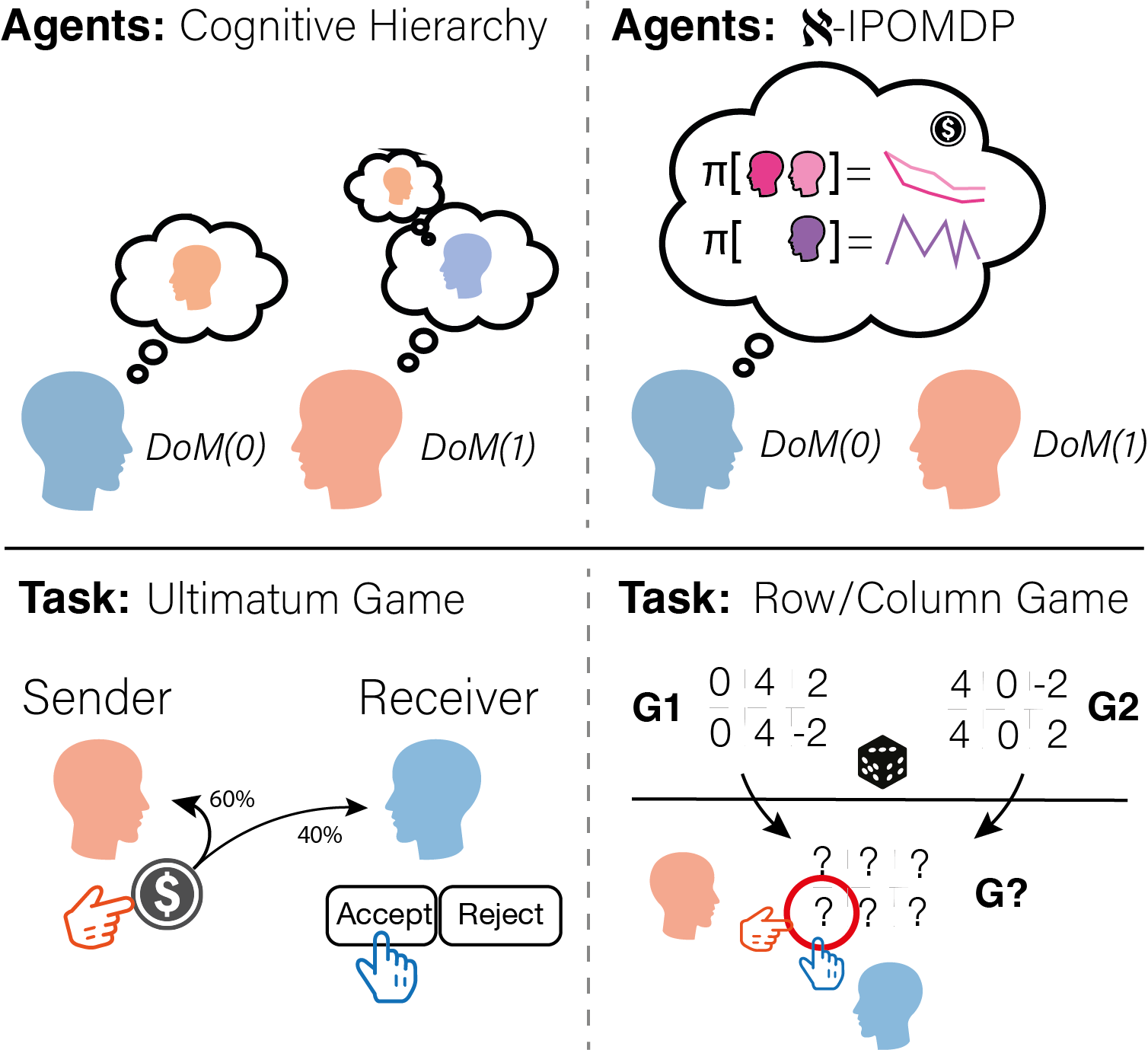}    
    \caption{\textbf{Paper overview}: \textbf{(Cognitive Hierarchy:)} We model agents with finite recursive opponent modelling with different Depth of Mentalising (DoM). In the classic model, the player at DoM(0) is at the mercy of the DoM(1) partner given that the DoM(0) player cannot form nested beliefs about their opponent. \textbf{($\aleph$-IPOMDP:)} The DoM(0) can overcome its recursive limitations by augmenting the classic model. We augment agents' inference processes with an anomaly detection mechanism that allows a self to detect deceptive others by matching expectations with observations. \textbf{(IUG:)} Agents with different degrees of DoM interact in the iterated ultimatum game (IUG). In the IUG, on each trial, a Sender offers a split of endowed money, and the Receiver decides whether to accept. If the Receiver decides to reject the offer, both players get 0. \textbf{(Row/column game:)} Agents interact in an iterated Bayesian zero-sum game. In this game, nature selects a payoff matrix ($G^1$ or $G^2$). Only the row player knows which payoff matrix is sampled and uses this information to their advantage. The column player makes inferences about the payoff matrix from the row player's behaviour.}
    \label{fig:introduction}
\end{figure}

In this work, we present a computational framework for multi-agent RL (MARL) called $\aleph$-IPOMDP. We augment the well-known IPOMDP approach  \citep{gmytrasiewicz_framework_2005} to allow agents to engage with unmodeled opponents. We first discuss how agents can use higher DoM to manipulate others, preying on the limited modelling capacity of their partners. We then present the main contribution of this work: a deception detection mechanism, the $\aleph$-mechanism, by which limited agents with shallow DoM use an off-policy counterfactual mechanism to detect anomalous behaviour, indicating that they are being deceived, and the OOB $\aleph$-policy, aimed at reasonably responding to the unknown opponent. We illustrate this mechanism in two game environments, mixed-motive and zero-sum Bayesian repeated games, to show how $\aleph$-IPOMDP agents can mitigate the advantages of agents with deeper recursive models, and successfully deter manipulation. Our work is relevant to multiple fields. To the MARL community, we show how agents with limited opponent modelling (for example, suffering from bounded rationality) can partially cope with more adequate opponents via anomaly detection and game-theoretic principles. To the cybersecurity community, we present a MARL masquerading detection use case which can be used to overcome learning adversarial attacks \citep{rosenberg2021adversarial}. To the psychology community, we provide an insights into how humans may use heuristics to detect deception when high DoM resources are not available, and why humans can avoid deception even in the absence of complex recursive reasoning. Given this framework, we can also show how human anomaly detection mechanisms may become maladaptively sensitive and overestimate deception in social environments without the need for over-mentalising, providing a context-appropriate model of suspicious or conspiratorial thinking. Lastly, there has recently been substantial interest in deception in AI  \citep{sarkadi_modelling_2019,sarkadi_evolution_2021,savas_deceptive_2022, masters2021characterising}, for which our work may serve as a blueprint for systems that regulate and possibly prevent AI agents from deceiving other AI or human agents.

\subsection{Previous Work}
The recursive structure of ToM implies that agents can not make correct inferences about other agents with DoM levels that are higher than their own. Modelling a more sophisticated agent would require a self to model itself from the perspective of another---outside of the DoM hierarchy--- and thus, at least in principle, to violate key logical principles \citep{pacuit_epistemic_2017}. Such a restriction is not unique to ToM-based models but is evident in general bounded rationality environments \citep{nachbar1996non}. The relation between deception and recursive ToM has been explored by \citet{oey2023designing}; in their work, however, agents are aware of the fact that others may deceive them and hence decide whether the others are lying or not. Nevertheless, humans take action under the belief that others are being deceptive, even in the absence of a fully fledged understanding of why. In our work, we consider the case in which agents cannot reason about the possibility of deceptive behaviour, but they can detect a deviation from \emph{expected} behaviour. This is somewhat analogous to off-the-equilibrium-path in Perfect Bayesian Equilibrium (PBE) --- certain behaviours (policies) should not appear in the interaction. If unexpected behaviour is observed, then it signals to the agent that something in the assumptions about behaviours is wrong. The problem of interacting with unmodeled others was introduced in ad-hoc teamwork \citep{stone2010ad, mirsky2022survey}. In a cooperative setting \citep{hu_off-belief_2021}, an agent needs to learn a policy (behaviour) to interact successfully with other agents. ToM allows successful interaction with others \citep{de_weerd_higher-order_2022}. It is, however, assumed that the others are properly modelled by the acting agent. Information-theoretic methods have previously been used for both deception modelling and deception detection. For example, \citet{kopp_information-theoretic_2018} introduced a deception modelling framework to study deceptive diffusion along with cooperation in populations exposed to fake news. Others have looked at detecting masquerading by inferring deception in the context of intrusion detection. For example, \citet{4455304} used the MDLCompress algorithm to identify intruders disguised as legitimate users. \citet{maguire2019seeing} suggests that humans apply a typical set-like mechanism to identify a non-random pattern. However, this is very specific to random behaviour. In this work, we advance this concept to explore several types of deviation away from expected behaviour. Behaviour-based Intrusion Detection Systems (IDS) methods were proposed by \citet{pannell_user_2010, peng_user_2016}. In these systems, the system administrator monitors the behaviour of users to decide and respond to malicious behaviour. However, unlike our proposed method, these systems often require labelled data, making them susceptible to an aware adversary who knows how to avoid detection. In the context of POMDP, several goal recognition methods have been proposed, for example \citep{ramirez_goal_withdate, le_guillarme_adversarial_2016}. While these methods assume that the observed agent may be malign, they assume that (a) the observer can invert the actions to make inferences about the malign intent, and (b) they use the likelihood of the observed behaviour to recognise this intent. Here, we show how likelihood-based inference is used \emph{against} the inferring agent, and propose a mechanism that detects deviation from expected behaviour, flagging malevolent agents without making inferences about their goals. \citet{yu_model-based_2021} explores how an agent might adapt to higher DoM opponents by learning the best response from experience. However, their mechanism lacks a model to detect when the opponent's DoM level exceeds the agent's DoM level, which is a necessary step to prompt the agent to retrain its model. On the other hand, \citep{piazza_limitations_2023} introduces a communicative MARL framework and shows that using ToM to defend against deceptive adversaries has several limitations that are determined by the multi-agent context in which communication is exchanged. For instance, the authors show that deceivers thrive in MARL settings where they can leverage biases such as static play, learning biases, and irrationality. Moreover, their model assumes that the victim can utilise their ToM to make inferences about the intention of the deceiver---distinguishing deceptive behaviour from non-deceptive (up to some limitation described in the paper). In this work we assume that the victim is deceived by an opponent with greater ToM capacity, depriving the victim of the ability to use its ToM to detect deception.

\subsection{Theory of Mind in Multi-agent RL}
In a partially observed single-agent RL problem (POMDP; \citep{kaelbling_planning_1998}), an agent is uncertain at time $t$ about the exact state of the environment, $s^t\in S$. It performs an action $a^t \in A$ and receives an observation $o^t \in O$ which depends on both of these: $p(o^t|a^t,s^{t})$. Assume that the dynamics of the environment are known to the agent (as we consider model-based RL) and that they are governed by a transition probability: $p(s^{t}|s^{t-1},a^{t-1})\equiv T(s^t,a^{t-1},s^{t-1})$. The agent makes an inference about the unknown state from the history, $h^{t-1}=\langle a^0,o^0,\dots, a^{t-1}, o^{t-1} \rangle$, in a Bayesian manner: $b(s^t) = p(s^t|h^{t-1})$. This can be expressed recursively as:
\begin{equation}
    b(s^t) = p(s^t|h^{t-1}) \propto P(o^{t-1}|a^{t-1},s^{t-1})\sum_{s^{t-1} \in S}T(s^t,a^{t-1},s^{t-1})b(s^{t-1})
    \label{eq:pomdp_belief_update}
\end{equation}

The agent's utility is a function of the environment and its actions: $u:A\times S \rightarrow \mathbb{R}$. An agent's goal is to maximize its long-term discounted utility: $\sum_{t=1}^T u^t\exp{[t\log(\gamma)]}$, where $\gamma \in [0,1]$ is the discount factor. In a POMDP, the optimal action-value function takes a particular action, and reports the sum of the expected immediate utility for that action at a belief state, and the expected long run discounted utility if the best possible action is subsequently taken at each subsequent step. It is defined recursively as:
\begin{align}
    &Q^*(a^t, b(s^t)) = E_{s^t \sim b(s^t)}\big[u^t(a^t, s^t) + \gamma  \sum_{o^t \in O}P(o^t|a^{t},s^{t}) \times
    \\
    \nonumber
    &\sum_{s^{t+1} \in S}T(s^{t+1},a^t,s^t)\underset{a^{t+1}}{\max}\{Q^*(a^{t+1}, b(s^{t+1}))\}\big]
\end{align}
The agent's policy is either a deterministic or stochastic function of these Q-values. One such stochastic policy is the SoftMax policy---a distribution over the actions with a known inverse temperature $\mathcal{T}$:
\begin{equation}
\pi(a^t|b(s^t)) = P(a^t|b(s^t)) \propto \exp{\frac{Q^*(a^t, b(s^t))}{\mathcal{T}}}
\label{eq:softmax_distribution}
\end{equation} In this work we assume that agents follow this policy.\footnote{Acknowledging the discrepancy with the definition of the optimal $Q^*$ values.}

In multi-agent RL (MARL), multiple RL agents interact with the same environment. We call the agents $\mu$ and $\nu$, and index their actions accordingly.\footnote{We discuss two agents, but the same principles apply to a larger number.} The interaction implies that the environment changes as a function of other agents' actions as well $p(s^t|a^{t-1}_\mu, a^{t-1}_\nu,s^{t-1})\equiv T(s^t,a^{t-1}_\mu, a^{t-1}_\nu,s^{t-1})$. Since the agent's utility is a function of the environmental state, it now becomes a function of other agents' actions, too. We highlight this dependency via reformulation of $\mu$'s utility: $u_\mu: A_\mu \times A_\nu \times S \rightarrow \mathbb{R}, u^t_\mu = f(a^t_\mu, a^t_\nu, s^t)$. Given this coupling, agents in MARL are motivated to predict the behaviour of others (\citet{wen_probabilistic_2019}). Driven by reward maximisation, and since its reward depends on $\nu$'s behaviour, $\mu$ may \emph{simulate} $\nu$'s reaction to $\mu$'s action and compute the expected utility of that action:
\begin{equation}
    E_{a^t_\nu}[u^t_\mu(a^t_\mu,a^t_\nu)] = \sum_{a^t_\nu} \hat P(a^t_\nu|a^t_\mu)u^t_\mu(a^t_\mu,a^t_\nu)
    \label{eq:simulate_opponent_behaviour}
\end{equation} where $\hat P(a^t_\nu|a^t_\mu)$ is $\mu$'s simulation of $\nu$. Thus, given a policy, $\pi_\mu$, agent $\mu$ may utilise this simulator to compute the discounted cumulative expected utility of this policy:
\begin{equation}
    E(u_\mu|\pi_\mu) = \sum_{t=0}^T E_{a^t_\nu}[u^t_\mu(a^t_\mu,a^t_\nu)]\exp{[t\log(\gamma)]}
    \label{eq:simulated_cumulative_discounted_utility}
\end{equation}
This is conceptually the mechanism governing the standard Nash equilibrium (NE). However, NE requires common knowledge (implying an arbitrarily deep level of nested knowledge) of the world and of others' behaviour \citep{chwe2013rational}. This assumption is easily revoked in multiple social interactions, where agents typically hold private and public information. Bayesian games \citep{zamir2020bayesian} model incomplete information encounters and are used to model social interaction. Formally, in a Bayesian game, agents do not have full information about the environment. This uncertainty may concern the state of the world (``Is there milk in the refrigerator?''), the intentions of other agents (``Friend or Foe?'') \citep{littman2001friend} or another's modus operandi (``how will Joe react to me doing this?''). Beliefs about unknown variables govern an agents' decision-making and are updated in a Bayesian manner, in a way similar to a single-agent POMDP. These beliefs may include beliefs about other agents' beliefs---a distribution over distributions \citep{da2024distributed}. Furthermore, agents can form beliefs about these recursive beliefs, i.e., a distribution over the distribution of distributions. This recursive reasoning is known as Theory-of-Mind (ToM). ToM is the ability to ascribe intentionality to others. This trait is considered one of the hallmarks of human cognition \citep{premack_does_1978}, and has been suggested to be vital to complex human behaviour, from effective communication \citep{goodman_pragmatic_2016, frank_predicting_2012} to deception \citep{alon2023dis, sarkadi_modelling_2019}. Due to the pivotal role ToM plays in the human capacity to interact socially, it has elicited interest from the AI community \citep{cuzzolin2020knowing, wang2021towards}. The IPOMDP framework \citep{gmytrasiewicz_framework_2005} combines ToM and POMDP, allowing us to model and solve these recursive beliefs models as a POMDP. While other models formalise recursive ToM (DoM), such as CHASE \citep{buergi_neural_2024}, hypergames \citep{bennett_hypergames_1980} or the aforementioned RSA model \citep{goodman_pragmatic_2016}, the IPOMDP generalises each of these frameworks into a flexible process that can be adapted to several social contexts. Naturally, while the principles presented in this work may be incorporated into any existing DoM model, we opted for the IPOMDP for its generality and wide use \citep{rusch_theory_2020,hula_model_2018,alon2023between}. In this model, an agent's \emph{type} \citep{harsanyi_games,westby_collective_2022}, $\theta_\mu$, is a combination of the agent's \emph{persona} (utility function, sensor capacity, etc.), denoted by $\psi_\mu$, and the agent's \emph{beliefs} about the world $b_\mu(\cdot)$: $\theta_\mu = \langle \psi_\mu, b_\mu(\cdot)\rangle$. It is assumed that agents are fully Bayesian, meaning that their uncertainty is epistemic and can be reduced to correctly identify the \emph{type} of agents with a lower DoM. Misplaced beliefs may only result from deceptive behaviour as explained in the following sections, and not from misplaced priors or insufficient support.

\paragraph{Inference with ToM} 
Beliefs can be limited to environmental uncertainty alone (reducing the problem to a POMDP) or they may also include beliefs from an intentional agent. The depth of belief recursion, $k \in [-1,\infty)$, is known as the agent's Depth of Mentalising (DoM). The Interactive State (IS) augments the concept of state in POMDP to account for the multi-variable nature of the problem (both environmental uncertainty and opponent uncertainty) $is_{\mu_k}^t = \langle s^t \times \theta_{\nu_{k-1}}^t \rangle$, where $\theta_{\nu_{k-1}}^t = \langle \psi_\nu, b_{\nu_{k-1}}(is_{\nu_{k-1}}^t) \rangle$ is the \emph{type} of the DoM$(k-1)$ agent---its persona and nested beliefs. This representation illustrates the recursive structure of ToM as it is possible to replace $is_{\nu_{k-1}}^t$ with $\langle s^t \times \langle \psi_\mu, b_{\mu_{k-2}}(is_{\mu_{k-2}}^t) \rangle \rangle$---revealing the hierarchical belief structure of ToM, in which $\mu_k$ reasons about $\nu_{k-1}$ reasoning about $\mu_{k-2}$ and so on. During the interaction, the acting agent $\mu$ observes the actions (either sequentially or simultaneously) of $\nu$ (that is $\mu$'s observations are $\nu$'s actions). In this work, assume that there is no environmental uncertainty, focusing on the strategic behaviour arising from uncertainty about the opponent  (we refer the reader to \citep{gmytrasiewicz_framework_2005} for a full description of belief update in IPOMDP). The resulting Bayesian updated beliefs about $\nu$'s type are:
\begin{equation}
b_{\mu_k}(is_{\mu_k}^t) = p(\theta^t_{\nu_{k-1}}|h^{t-1}) \propto  \sum_{\theta^{t-1}_{\nu_{k-1}} \in \Theta_{\nu_{k-1}}}P(a_{\nu}^{t-1}|\theta^{t-1}_{\nu_{k-1}}, a_{\mu}^{t-1})p(\theta^{t-1}_{\nu_{k-1}}|h^{t-2}) 
\label{eq:marl_recursive_belief_update}
\end{equation}
Since $\mu$'s actions affect $\nu$'s beliefs, but $\nu$'s persona ($\psi_\nu$) is assumed immutable, we expand Eq.~\ref{eq:marl_recursive_belief_update} to distinguish between inference about temporal changes to $\nu$'s belief (potentially about $\mu$'s type) and inference about $\nu$'s unalterable persona:
\begin{equation}    
p(\theta^t_{\nu_{k-1}}|h^{t-1}) = p(\langle b_{\nu_{k-1}}^{t}, \psi_\nu\rangle|h^{t-1}) \propto  \sum_{\psi_\nu \in \Psi{\nu}}\big( P(a_{\nu}^{t-1}|\langle b_{\nu_{k-1}}^{t-1}, \psi_\nu\rangle, a_{\mu}^{t-1})\times P(b_{\nu_{k-1}}^{t}|b_{\nu_{k-1}}^{t-1},a_{\mu}^{t-1})\big) p(\theta^{t-1}_{\nu_{k-1}}|h^{t-2}) 
\label{eq:detailed_marl_recursive_belief_update}
\end{equation}
\paragraph{The Cognitive Hierarchy}
At the core of recursive models are subintentional agents (sometimes referred to as DoM$(-1)$). These agents follow reactive, typically model-free and myopic policies, treating other agents' actions as environmental features rather than strategic choices. Their optimal policy depends only on their own policy and the interaction history: 
\begin{align}
    \pi_{\mu_{-1}}^t(\psi_{\mu}, a_\nu^{t-1}) &= P(a^t_\mu| \psi_{\mu}, a_\nu^{t-1})
    \label{eq:dom_m1_policy}
\intertext{The DoM$(0)$ models other agents as having DoM$(-1)$, implying that these agents model the opponent and environment separately. Plugging Eq.~\ref{eq:dom_m1_policy} into Eq.~\ref{eq:marl_recursive_belief_update} yields the DoM$(0)$ belief update, which is a simple Bayesian IRL \citep{ramachandran_bayesian_withdate}:}
    b_{\nu_0}(\theta_{\nu_0}^t) &= p(\psi_{\mu}|h^{t-1}) \propto \sum_{\psi_{\mu} \in \Theta_{\mu_{-1}}}P(a_\mu^{t-1}|\psi_{\mu}, a_\nu^{t-1})p(\psi_{\mu}|h^{t-2})    
    \label{eq:dom_zero_belief_update}
\intertext{Using these beliefs the DoM$(0)$ computes the state-action values (Q-values):}
    Q^*_{\nu_0}(a_\nu^t,\theta_{\nu_0}^t) &= E_{a_\mu^t \sim \pi_{\mu_{-1}}(\psi_{\mu}, a_\nu^{t-1})}\big[u_\nu^t(a_\nu^t, a_\mu^t) + \max_{a^{t+1}_\nu} [Q^*_{\nu_0}(a_\nu^{t+1},\theta_{\nu_0}^{t+1})] \big]
    \label{eq:q_values_of_dom_zero}
\intertext{where $u_\nu^t(a_\nu^t, a_\mu^t)$ is $\nu$'s utility at time $t$. The DoM$(1)$ models others as DoM$(0)$ agents, making inferences about their beliefs as well:}
    b_{\mu_1}(\theta_{\mu_1}^t) &= p(\psi_{\nu} \times b_{\nu_0}(\theta_{\nu_0}^t)|h^{t-1}) \propto P(a_\nu^{t-1}|\psi_{\nu}, b_{\nu_0}(\theta_{\nu_0}^{t-1}), a_\mu^{t-1})\times
    \label{eq:dom_one_belief_update}    
    \\
    \nonumber
    & \hspace*{1.7in}P(b_{\nu_0}(\theta_{\nu_0}^{t})|b_{\nu_0}(\theta_{\nu_0}^{t-1}), a_\mu^{t-1})p(\psi_{\nu}, b_{\nu_0}(\theta_{\nu_0}^{t-1})|h^{t-2})   
\end{align}
This trait allows the DoM$(1)$ to reason about changes in the beliefs of DoM$(0)$ and use this ability to its benefit, as presented next. Similarly to the DoM$(0)$, the DoM$(1)$ computes the $Q$-values of its actions and acts based on these values.

\subsection{Planning with ToM}

Theory of Mind allows the acting agent to simulate other agents, taking their perspective (Eq.~\ref{eq:simulate_opponent_behaviour}). Formally, this is related to the concept of sequential rationality in extensive-form Bayesian games \citep{harsanyi_games, fudenberg_perfect_1991}. In the current work, the acting agent maintains a belief system over the opponent's hidden 'types' --- represented by their DoM level and persona --- and computes optimal policies modulo this mental model. This capacity empowers agents with high DoM to shape the behaviour (\citep{jaques_social_2019, kim2022influencing}) of lower DoM agents to their benefit. We formally describe this process before illustrating how the utility function of the acting agent is integrated into its policy computation. Using its nested mental model of $\nu$, denoted $\hat \theta_{\nu_{k-1}}$, the acting agent $\mu_k$ computes $\nu$'s optimal policy, given its model of $\nu$'s belief and persona:
\begin{equation}
    \hat\pi^t_{\nu_{k-1}}(\hat \theta^t_{\nu_{k-1}}) = p(a^t_\nu|\hat b_{\nu_{k-1}}(\hat{\hat{\theta}}^t_{\mu_{k-2}}),\psi_\nu)        
    \label{eq:simulated_opponent_behaviour}
\end{equation}  
where $\hat\pi^t_{\nu_{k-1}}(\hat{\cdot})$ denotes the simulated policy of a nested opponent model and $\hat b_{\nu_{k-1}}(\hat{\hat{\theta}}^t_{\mu_{k-2}})$ denoted the nested belief about a nested self model---the belief $\mu_k$ ascribes to $\nu_{k-1}$ about $\mu$, as modelled by $\nu$. From Eq.~\ref{eq:simulated_opponent_behaviour} $\mu$ computes a distribution of future actions conditioned on a given policy:

\begin{equation}
p(a^{t+1:T}_{\nu}|\hat{\theta}^t_{\nu_{k-1}} ,\pi^t_{\mu_k}) = \prod_{i=1}^{T-t}\sum_{a_\mu^{t+i-1}}
p(a_\nu^{t+i}|\hat\pi^{t+i}_{\nu_{k-1}}(\hat\theta^{t+i}_{\nu_{k-1}}))p(\hat b_{\nu_{k-1}}(\hat{\hat\theta}^{t+i-1}_{\mu_{k-2}})|a_\mu^{t+i-1})P(a_\mu^{t+i-1}|\pi_{\mu_k}^{t+i-1})
\label{eq:simulation_of_nested_model_policy}
\end{equation} where $\pi_{\mu_k}^{t}$ is $\mu$'s revised policy after observing $a_\nu^{t-1}$ and updating its beliefs about $\hat{\theta}^{t}_{\nu_{k-1}}$. The interplay between $\mu$'s utility and $\nu$'s behaviour affects $\mu$'s value function. The \emph{value} function of a policy $\pi_{\mu_k}$ \citep{sutton_reinforcement_2018} is defined as the expected cumulative utility of $\mu_k$ if it samples its actions from it ($a_\mu^{t} \sim \pi^t_{\mu_k}(\psi_\mu, b_{\mu_k}(\theta^{t}_{\nu_{k-1}}))$):
\begin{equation}
    V^{\pi^t_{\mu_k}}(b_{\mu_k}(\theta^t_{\nu_{k-1}})) = E_{\pi^t_{\mu_k}}[E_{a^t_\nu \sim \hat \pi^t_{\nu_{k-1}}(\psi_\nu, b_{\nu_{k-1}}(\theta^{t}_{\mu_{k-2}}))}\big [u_\mu(a_\mu^t, a_\nu^t) + \gamma V^{\pi^{t+1}_{\mu_k}}(b_{\mu_k}(\theta^{t+1}_{\nu_{k-1}}))]\big]
    \label{eq:value_function_mu_marl}
\end{equation}

Changes in the policy are propagated to the value function through two components - an \emph{immediate} effect and a \emph{gradual} effect caused by changes in $\nu$'s behaviour. We illustrate these effects to highlight the role of ToM in optimal (at least from the DoM($k$) agent's perspective) policy
(the full computation is presented in App. ~\ref{app:derivation of value gradient by policy}):
\begin{equation}
\frac{\partial V^{\pi^t_{\mu_k}}(b_{\mu_k}(\theta^t_{\nu_{k-1}}))}{\partial \pi^t_{\mu_k}} = \frac{\partial E_{\pi^t_{\mu_k}}[E_{\pi^t_{\nu_{k-1}}}[u^t_\mu(a^{t}_\mu,a^{t}_\nu)]]}{\partial \pi^t_{\mu_k}} + \gamma \frac{\partial E_{\pi^t_{\mu_k}}[E_{\pi^t_{\nu_{k-1}}}[V^{\pi^{t+1}_{\mu_k}}(b_{\mu_k}(\theta^{t+1}_{\nu_{k-1}}))]]}{\partial \pi^{t}_{\mu_k}}
\label{eq:policy_effect_on_value_in_marl}    
\end{equation}    
The immediate effect (first RHS argument of Eq. ~\ref{eq:policy_effect_on_value_in_marl}) measures the change of $\mu$'s current utility as a function of its action (assuming simultaneous actions, but this can be adapted to sequential actions, too):
\begin{equation}
\frac{\partial E_{\pi^t_{\mu_k}}[E_{\pi^t_{\nu_{k-1}}}[u^t_\mu(a^{t}_\mu,a^{t}_\nu)]]}{\partial \pi^t_{\mu_k}} = 
\sum_{a_\mu^t}\frac{\partial P(a^t_\mu|\pi^t_{\mu_k})}{\partial \pi^t_{\mu_k}}\bar{u}(a^t_\mu)
\label{eq:policy_immediate_effect}
\end{equation}
Where $\bar{u}(a^t_\mu) = E_{\pi^t_{\nu_{k-1}}}[u^t_\mu(a^{t}_\mu,a^{t}_\nu)]$ is the expected utility for $\mu$ from playing action $a^t_\mu$ averaged over $\nu$'s policy.

The gradual effect ($t+1$) describes the long-term effect on $\mu$'s value, stemming from $\nu$'s adapted behaviour to $\mu$'s action (\citep{siu_towards_2022, na_humans_2021}):
\begin{align}
    &\frac{\partial E_{\pi^t_{\mu_k}}[E_{\pi^t_{\nu_{k-1}}}[V^{\pi^{t+1}_{\mu_k}}(b_{\mu_k}(\theta^{t+1}_{\nu_{k-1}}))]]}{\partial \pi^{t}_{\mu_k}} = 
    \label{eq:action_gradual_effect}    
    \\
    \nonumber
     &\sum_{a^t_\nu}P(a^t_\nu|\pi^t_{\nu_{k-1}})\big[\sum_{a^t_\mu}\frac{\partial P(a^t_\mu|\pi^t_{\mu_k})}{\partial \pi^{t}_{\mu_k}}\times V^{\pi^{t+1}_{\mu_k}}(b_{\mu_k}(\theta^{t+1}_{\nu_{k-1}})) + 
    P(a^t_\mu|\pi^t_{\mu_k})\times 
    \frac{\partial V^{\pi^{t+1}_{\mu_k}}(b_{\mu_k}(\theta^{t+1}_{\nu_{k-1}}))}{\partial \pi^{t}_{\mu_k}}\big]
\end{align}
The term $\frac{\partial P(a^t_\mu|\pi^t_{\mu_k})}{\partial \pi^{t}_{\mu_k}}V^{\pi^{t+1}_{\mu_k}}(b_{\mu_k}(\theta^{t+1}_{\nu_{k-1}}))$ relates to \emph{policy gradient} \citep{sutton_policy_1999}. The second term corresponds to \emph{opponent shaping} \citep{foerster_learning_2018}. 
Unlike the work of \citet{foerster_learning_2018}, the ToM opponent shaping takes an \emph{indirect} path, whereby the shaping agent induces behavioural change by affecting others' beliefs. This agency is expressed through the following equation (see App. \ref{app:derivation of value gradient by policy} for full derivation):
\begin{equation}
    \frac{\partial V^{\pi^{t+1}_{\mu_k}}(b_{\mu_k}(\theta^{t+1}_{\nu_{k-1}}))}{\partial \pi^{t}_{\mu_k}} = 
    \frac{\partial V^{\pi^{t+1}_{\mu_k}}(b_{\mu_k}(\theta^{t+1}_{\nu_{k-1}}))}{\partial \pi^{t+1}_{\nu_{k-1}}}
    \frac{\partial \pi^{t+1}_{\nu_{k-1}}}{\partial b_{\nu_{k-1}}(\theta^{t+1}_{\mu_{k-2}})}
    \frac{\partial b_{\nu_{k-1}}(\theta^{t+1}_{\mu_{k-2}})}{\partial \pi^{t}_{\mu_k}}    
\end{equation}
First, $\mu$'s action changes $\nu$'s belief: $\frac{\partial b_{\nu_{k-1}}(\theta^{t+1}_{\mu_{k-2}})}{\partial \pi^{t}_{\mu_k}}$, computed via derivation of Equation~\ref{eq:marl_recursive_belief_update}. For example, if $\nu$'s beliefs are parametric, this change is computed via the \emph{influence} function \citep{koh2017understanding}. Next, the changes in $\nu$'s belief affect its Q-value computation and hence its policy $\frac{\partial \pi^{t+1}_{\nu_{k-1}}}{\partial b_{\nu_{k-1}}(\theta^{t+1}_{\mu_{k-2}})}$. Moreover, the long term effect of shaping $\nu$'s beliefs in favour of $\mu$ is propagated into future steps through $\nu$'s updated beliefs (as illustrated in Sec. ~\ref{subsec:mixed motive games}). In a cooperative game, shaping is often the revelation of unknown information or a nudge, aimed to improve others utility \citep{jaques_social_2019}, as it is assumed that the agents' goal is shared and therefore there is no incentive for $\mu$ to hide or falsify information \citep{devaine_social_2014}. However, this is not the case in non-cooperative games, where often agents can gain from disclosing information or providing signals that cause the observer to form \emph{false beliefs} \citep{alon2023between,alon2023dis}; this concept is presented in the next section.

\section{Deception with ToM}
\label{sec:deception_with_tom}

Deception is defined as the deliberate causation of a false belief in the mind of a target with an ulterior motive \citep{sarkadi_modelling_2019, kopp_information-theoretic_2018, ward_defining_2023}. However, assessing the success of deception is non-trivial; there is often no guarantee that a victim's action was induced by the intended false belief or if the victim acted independently \citep{masters_deceptive_2017}. We formalize and rigorously operationalize this in our framework. We propose an axiomatic construction of deception based on sequential rationality and Depth of Mind (DoM). We define a deceptive policy $\pi_\mu^\dagger$ through four necessary conditions: Incentive Compatibility, Epistemic Manipulation, Behavioural Causality, and Cognitive Dominance.

\paragraph{Axiom 1: Incentive Compatibility (The Ulterior Goal)}
\label{par:axiom_1}
A rational agent will only engage in deception if it yields a higher expected utility than honest behaviour. Formally, let $\pi_\mu^\dagger$ denote a policy that knowingly aims to install a false belief. The agent $\mu$ will only choose $\pi_\mu^\dagger$ if it is at least as valuable as any other policy:
\begin{equation}    
    V^{\pi_\mu^\dagger}(b_{\mu_k}(\theta_{\nu_{k-1}})) \geq V^{\pi_\mu}(b_{\mu_k}(\theta_{\nu_{k-1}}))
    \label{eq:value_of_deceptive_policy_deceiver}
\end{equation} 
This inequality satisfies the motivation established in Eq.~\ref{eq:policy_effect_on_value_in_marl}: $\mu$ is incentivised to ``affect'' $\nu$ specifically because the resulting behaviour generates excess utility.

\paragraph{Axiom 2: Epistemic Manipulation (False Beliefs)}
Deception requires the transmission of a signal intended to induce a divergence between the victim's belief and the deceiver's true parameters. By false belief, we mean a belief that $\mu$ believes (knows) is false \cite{ward_defining_2023}, effectively targeting the victim's \emph{perception} \citep{eger_practical_2017}. A policy $\pi_\mu^\dagger$ for a deceiver of type $\theta_\mu^\dagger = \langle \psi_{\mu^\dagger}, b_{\mu^\dagger}(\cdot)\rangle$ installs \emph{false beliefs} if, during a deceptive phase $\bm{T_D} = \{t_1,\dots,t_D\} \subseteq \{0,\cdots,T\}$, the victim assigns higher probability to an incorrect type $\theta^\star_\mu$ than the true type $\theta^\dagger_\mu$:
\begin{equation}
    \exists \bm{T_D}: \forall t \in \bm{T_D}: P_\nu(\theta^\star_\mu|h^t) \geq P_\nu(\theta^\dagger_\mu|h^t)
    \label{eq:false_beliefs}
\end{equation} 
Given the cognitive hierarchy prohibition on modelling higher DoM-level agents' beliefs, we decompose Eq. 20 to highlight two aspects of the deceptive policy: installing false beliefs about the \emph{persona} $\psi$, and manipulation of nested beliefs. At any DoM level, the deceiver manipulates the victim to falsely believe that its persona $\psi_{\mu}$ is $\psi_{\mu^\star}$ rather than $\psi_{\mu^\dagger}$:
\begin{equation}
    \exists \bm{T_D}: \forall t \in \bm{T_D}: P_\nu(\psi_{\mu^\star}|h^t) \geq P_\nu(\psi_{\mu^\dagger}|h^t)
    \label{eq:false_persona_beliefs}
\end{equation}
For example, a deceptive foe (higher DoM) may portray itself as a friend \citep{adhikari2021telling}, satisfying this axiom by manipulating the probability mass over the persona set. Recursing through higher levels ($k \geq 2$), we use Eq. ~\ref{eq:false_persona_beliefs} to define \emph{false nested beliefs}. Crucially, by definition, the victim (DoM($k$)) cannot fully represent the deceiver's beliefs, as they include ($k+1$) beliefs, implying that the victim's beliefs about the deceiver's \emph{type} are wrong. Nonetheless, since both agents share at least $(k-2)$ beliefs (for example, if $k = 3$, they share both level $1$ and level $0$ beliefs) --- the manipulation of these shared beliefs is part of the deception. That is, as part of the ploy, the higher DoM deceiver can manipulate the victim's beliefs about $\mu$'s (the deceiver) beliefs about $\nu$ (the victim). For example, the deceiver may aim to not only cause the victim falsely believe that they are friendly, but also to induce false (nested) beliefs that they themselves ($\mu$) believe that $\nu$ is a friend too. Formally, let $p^\dagger_\mu(\psi_\nu|h^t)$ denote $\mu$'s \emph{true} belief about $\nu$'s persona. Similarly, we define the \emph{false} (or wrong) belief as $p^\star_\mu(\psi_\nu|h^t)$. Using these notations, and the $\hat{\cdot}$ notation for estimated nested beliefs, we define $\nu$'s \emph{false nested belief} about $\mu$'s beliefs as:
\begin{equation}
    \exists \bm{T_D}: \forall t \in \bm{T_D}: P_\nu(\hat{p}_\mu^\star(\psi_{\nu}|h^t)) \geq P_\nu(\hat{p}_\mu^\dagger(\psi_{\nu}|h^t))
    \label{eq:false_nested_beliefs}
\end{equation} 

\paragraph{Axiom 3: Behavioural Causality (Regret)}
It is not sufficient for $\nu$ to merely hold a false belief; that belief must ``tip the scales'' \citep{sarkadi_deception_2021} to induce a behaviour that is suboptimal for the victim but beneficial for the deceiver. We measure this via \emph{regret} \citep{nisan_learning_2007,jin_regret_2018}, defined here as the difference between the victim's chosen action (based on false belief $b_\nu^{t,\dagger}$) and the optimal action they would have taken had they possessed the cognitive capacity (DoM($k+1$)) to see through the bluff:
\begin{equation}
    Reg_{\nu_{k-1}}^t = E_{a_\nu^t \sim \pi^t_{\nu_{\bm{k-1}}}(b_\nu^{t,\dagger})} u^t_\nu(a_\mu^t, a_\nu^t) - E_{a_\nu^t \sim \pi^t_{\nu_{\bm{k+1}}}(b_\nu^{t})} u^t_\nu(a_\mu^t, a_\nu^t)
    \label{eq:regret_of_deceptive_policy}
\end{equation} 
Since the victim cannot compute this counterfactual, we approximate the discrepancy by comparing the realized reward against the victim's expected reward under their (manipulated) belief. Using Eq.~\ref{eq:simulated_opponent_behaviour}, the victim's expected reward for a given belief state is:
\begin{equation}
    E(\hat r_\nu^t) = \sum_{\hat\theta_{\mu_{k-2}}}\hat r_\nu^t(\hat\theta_{\mu_{k-2}}) P_{\nu_{k-1}}^{t-1}(\hat\theta_{\mu_{k-2}})
    \label{eq:expected_reward_for_given_belief}
\end{equation}
The measurable estimator for the success of the deception is thus:
\begin{equation}
\hat{Reg}^t = r_\nu^t - E(\hat r_\nu^t)
\label{eq:cumulative_reward_difference}
\end{equation}
In zero-sum games, a negative $\hat{Reg}^t$ for the victim corresponds to the exact advantage gained by the deceiver.

\paragraph{Axiom 4: Cognitive Dominance (Avoiding Detection)}
Finally, for deception to persist, the victim must remain unaware of the manipulation. This necessitates a cognitive asymmetry where $DoM(\mu) > DoM(\nu)$. Because $\nu$ is limited to making inferences about lower-level DoM agents, it cannot straightforwardly interpret the deceiver's actions as deceptive without violating logical principles of self-reflection \citep{pacuit_epistemic_2017}. We conclude that ToM is the necessary mechanism to satisfy these axioms. It enables the simulation required for Axiom 2, the policy optimization for Axiom 1, and the exploitation of the cognitive gap defined in Axiom 4 to induce the regret defined in Axiom 3.

\subsection{Illustration in Bayesian Games}
\label{sec:illustration_in_bayesian_games}
We illustrate deceptive behaviour in a mixed-motive and a zero-sum game. Since the main results and findings are similar, we present the mixed-motive game results in the main body and the zero-sum game results in Appendix~\ref{app:zero_sum_description}. In both cases, we illustrate how higher DoM agents utilise nested models to simulate lower DoM agents, and how this nesting is used to maximise reward. We discuss the details of the deception concerning Eq.~\ref{eq:policy_effect_on_value_in_marl} and show how the policy balances short- and long-term rewards. We conclude that ToM-based deception follows a pattern of installing false beliefs, followed by the execution of undetected strategic behaviour.

\subsubsection{Mixed-Motive Game}
\label{subsec:mixed motive games}
\begin{figure}[ht]
    \centering
    \includegraphics[scale=.08]{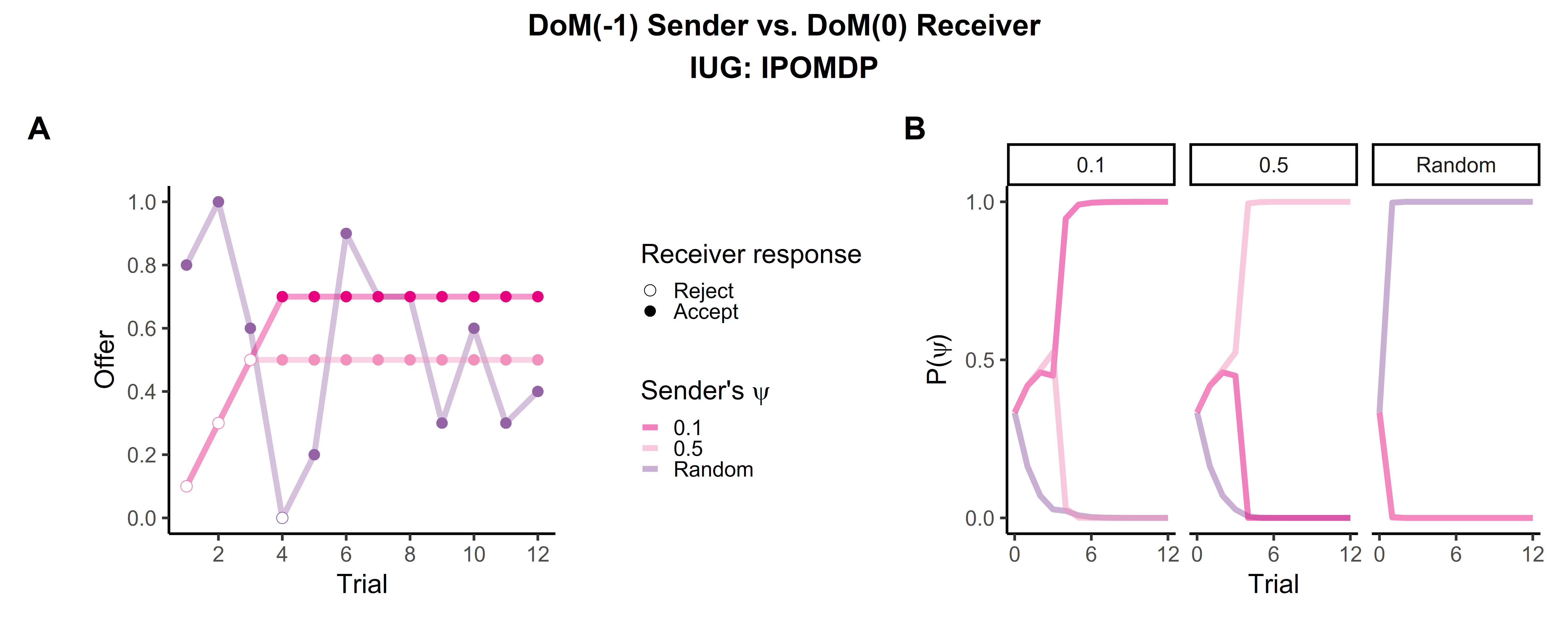}    
    \caption{\textbf{DoM$(0)$ vs DoM$(-1)$ in IUG IPOMDP}: \textbf{(A)} The points show offers from the sender to the receiver over all 12 trials, coloured by sender behavioural type (random or utility). Points are shown in white if the receiver rejects the offer. A DoM$(0)$ receiver quickly infers from the initial offers the type of the DoM$(-1)$ sender. The DoM$(-1)$ utility sender's first offer tells it apart from the random sender as its initial offer is always close to $0$ (up to random noise). The DoM$(0)$ policy is a function of its updated beliefs \textbf{(B)}. Updated belief probabilities of the receiver when playing with different random or utility senders. DoM$(0)$ receivers are well tuned to detect which type of sender they are partnered with. When engaging with a threshold DoM$(-1)$ sender, the receiver rejects the offers until the sender is unwilling to ``improve'' its offers, which also corresponds to the certainty of its beliefs}
    \label{fig:iug_strategic_behaviour_illustration}
\end{figure}

The iterated ultimatum game (IUG) \citep{camerer2011behavioral,alon2023between} (illustrated in Fig ~\ref{fig:introduction}(Ultimatum Game)); is a repeated Bayesian mixed-motive game \citep{alon2023between}. Briefly, the IUG is played between two agents---a sender and a receiver. We will consider how the sender might try to deceive the receiver, and so designate them as $\mu$ and $\nu$ respectively, following the convention established above. On each trial $t$ of the game, the sender gets an endowment of 1 monetary unit and offers a (for convenience, discretized) percentage of this to the receiver: $a^t_\mu \in \{0,0.1,0.2\ldots,1\}, t\in [1,T]$ (we used $T$=12). If the receiver \emph{accepts} the offer $(a^t_\nu = 1)$, the receiver gets a reward of $r^t_\nu = a^t_{\mu}$ and the sender a reward of $r^t_\mu = 1-a^t_{\mu}$. Alternatively, the receiver can \emph{reject} the offer $(a^t_\nu = 0)$, in which case both players receive nothing. In this game, agents need to compromise (or at least consider the desires of each other) in order to maximise the reward that they can achieve---this makes the IUG a useful testbed for assessing strategic mentalising. We simulated senders with two DoM levels: $k\in [-1,1]$ interacting with a DoM$(0)$ receiver. The DoM($-1$) senders came in three behavioural types: one emits uniformly random offers (and so is not endowed with  Q-values); the other two 
have utility functions that are characterized by a threshold, $\psi_\mu \in \{0.1, 0.5\}$: $u^t_\mu(a^t_{\mu},a^t_{\nu}, \theta_\mu) = (1- a^t_{\mu} - \psi_\mu) \cdot a^t_{\nu}$. We call these senders the threshold senders.  The threshold senders compute Q-values based on their models of the world and select their actions via the SoftMax policy (Eq.~\ref{eq:softmax_distribution}). The DoM$(-1)$ threshold senders compute the Q-values in the following way---they maintain lower and upper bounds on the viable offer set, $L^t, U^t$. These bounds are updated:
\begin{align}
    L^t &= L^{t-1}\cdot a_\nu^{t-1} + a_\mu^{t-1}\cdot (1-a_\nu^{t-1}) 
    \\
    U^t &= U^{t-1}\cdot (1-a_\nu^{t-1}) + a_\mu^{t-1}\cdot (a_\nu^{t-1}) 
    \label{eq:dom_m1_bounds}
\end{align} with $L^0 = 0$ and $U^0=1$. In this work, we assume that the DoM$(-1)$ agents are extremely myopic, reducing the Q-values to the immediate reward \citep{alon2023between}. In turn, these senders' Q-values (in fact, degenerate Q-values) are simply the immediate utility from every action in the range $a_\mu^t \in (L^t, U^t]$:
\begin{equation}
    Q_{\mu_{-1}}^t(a_\mu^t;\psi_\mu) = u_\mu^t(a_\mu^t, \psi_\mu)
    \label{eq:dom_minus_one_q_values_iug}
\end{equation} and not computing the values for actions outside of this interval (narrowing the possible actions after each iteration).
This assumption can be revoked in future work. The sender will select an offer (via SoftMax policy) from the set of potential offer based on the offer's Q-value. We model these agents as rational agents, meaning they will not engage in a losing interaction. Lacking the option to quit, if the lower bound $L^t$ reaches the sender's threshold $\psi_\mu$ we assume that the sender will not update the bounds and will offer a partition that yields it a zero reward.
Since these agents are assumed to lack an opponent model and are often modelled \citep{gmytrasiewicz_framework_2005} as having a uniform distribution over future events, we expect the results to hold. The DoM$(0)$ receiver infers the behavioural type of the DoM$(-1)$ sender from their action as in Eq.~\ref{eq:dom_zero_belief_update}. As described above, we assume that the DoM$(0)$ receiver has full knowledge of the potential types of senders ($\Theta_{\mu_{-1}}$) and its uncertainty is limited to which realization is sampled $\psi_{\mu_{-1}} \in \Theta_{\mu_{-1}}$. We leave uncertainty about $\Theta_\mu$ to future work.
Given the updated beliefs, they compute the Q-values (Eq.~\ref{eq:q_values_of_dom_zero}), via Expectimax search \citep{russell2010artificial} and select an action via a SoftMax policy (Eq.~\ref{eq:softmax_distribution}) with temperature ($\mathcal{T} = 0.1$; which is common knowledge). Lastly, the DoM$(1)$ sender simulates the DoM$(0)$ receiver's beliefs and resulting actions during the computation of its Q-values using the IPOMCP algorithm \citep{hula_monte_2015}, with full planning horizon. As hypothesised, agents with high DoM take advantage of those with lower DoM. The complexity of this manipulation rises with the agents' level. The DoM$(0)$ receiver infers the type of the DoM$(-1)$ sender and uses its actions to manipulate the behaviour of the sender (if possible). Specifically, if the receiver believes they are partnered with an intentional (non-random) sender, they will try to influence the sender to offer more via strategic rejection (as illustrated in Fig.~\ref{fig:iug_strategic_behaviour_illustration}, illustrating one simulation). This will continue until either the sender's threshold is met or until the offers are sufficient given the opportunity cost implied by the planning horizon. Note that this type of behaviour is deceptive according to our definition, as the receiver uses its model of the sender to cause the sender to improve its offers, which improves the receiver's utility, by planting ``false-beliefs'' in the DoM$(-1)$ victim's ``mind''. On the other hand, if a receiver believes they are partnered with a random sender, then, since the random sender affords no agency, the optimal policy for the receiver is to accept any offer.

\begin{figure}[htbp]
    \centering
    \includegraphics[scale=.06]{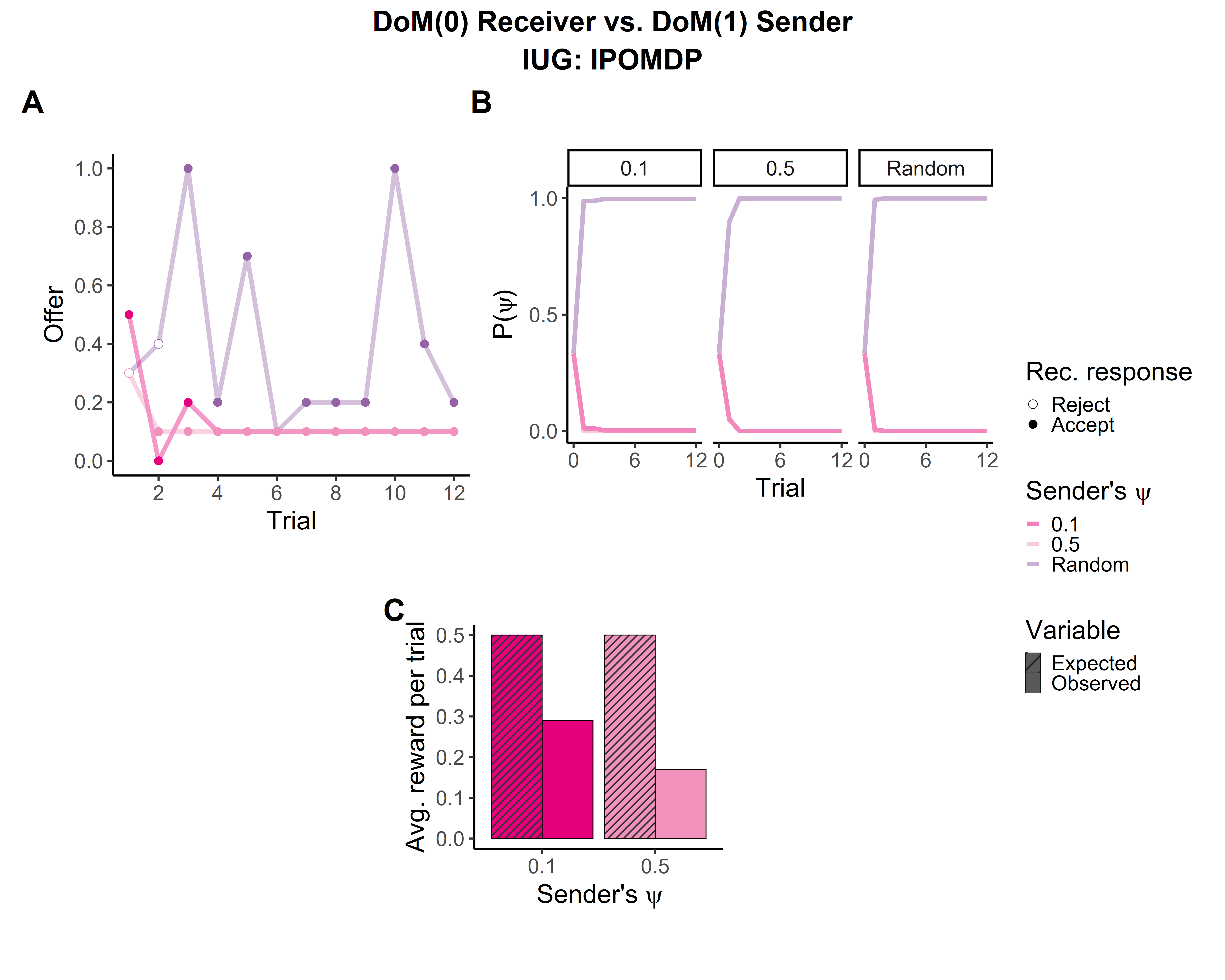}        
    \caption{\textbf{Illustration of deception in IUG}: \textbf{(A)} Points show offers from the sender to the receiver over all 12 trials, coloured by sender behavioural type (random or utility). Points are shaded white if the receiver rejects the offer. The DoM$(1)$ acts in a deceptive way to masquerade itself as a random sender, hacking the DoM$(0)$ Bayesian IRL. It starts with a relatively high first offer, and then decreases sharply. \textbf{(B)} Updated belief probabilities of the receiver when playing with different random or utility senders. DoM$(0)$ receivers are poorly tuned to detect the type of DoM$(1)$ sender with which they are partnered, mistaking all actions as if they came from the random sender. This stratagem employed by the sender exploits the pitfall of Bayesian inference used by the DoM$(0)$---the likelihood that any offer sequence is equal for the random sender. \textbf{(C)} Comparing the expected reward from the DoM$(0)$ receiver's perspective $E(\hat r_\nu^t)$ (striped bars) to the observed reward (non striped bars) each trial, This measures the advantage of the deceiver's policy, reflecting the deceiver's ability to increase its reward at the expense of the victim.}
    \label{fig:mixed_motive_deception_illustration}
\end{figure}

The DoM$(1)$ sender's Q-values are computed in a similar manner to Eq.~\ref{eq:q_values_of_dom_zero}:
\begin{align}
    Q^*_{\mu_1}(a_\mu^t,\theta_{\mu_1}^t) &= E_{a_\nu^t \sim \pi_{\nu_{0}}(\psi_{\nu}, b_{\nu_0}(\psi_{\mu_{-1}}))}\big [u_\mu^t(a_\mu^t, a_\nu^t) + \gamma\max_{a^{t+1}_\mu} [Q^*_{\mu_1}(a_\mu^{t+1},\theta_{\mu_1}^{t+1})] \big]
    \label{eq:dom_one_sender_q_values}
\end{align} where, as explained in Section \ref{sec:deception_with_tom}, the actions of the DoM$(1)$ sender affect not only its immediate reward, but also the long term reward via the belief manipulation process. The DoM$(1)$ sender sets out to deceive the DoM$(0)$ receiver based on the former's behavioural pattern. Using its capacity to simulate this behaviour fully, the DoM$(1)$ sender acts deceptively by masquerading as a random sender, preying on the lack of agency that the DoM$(0)$ has over the random sender. This policy is depicted in Fig.~\ref{fig:mixed_motive_deception_illustration}. 

The sender begins by making a relatively high offer (Fig.~\ref{fig:mixed_motive_deception_illustration}(A)). This offer is very unlikely for the threshold DoM$(-1)$ senders, hence the belief update of the DoM$(0)$ receiver strongly favours the random sender hypothesis (Fig.~\ref{fig:mixed_motive_deception_illustration}(B)) (Axiom 2). Once the receiver's beliefs are misplaced, the sender takes advantage of the statistical nature of the random sender policy---every offer has the same likelihood: $\frac{1}{|\mathcal{A}|}$. This allows the sender to reduce their subsequent offers substantially (Axiom 1), while avoiding ``detection'' (Axiom 4). Thus, the DoM$(1)$ policy is deceptive policy according to our definition. We note that any deceptive schema depends on the types the deceiver can masquerade to (Eq. \ref{eq:false_beliefs}). In particular, the existence of the random sender enables the DoM$(1)$ sender to act in a way that benefits it and installs false beliefs in the receiver's mind. Even with the lack of a noisy agent, as long as there are other types for which the behaviour of the victim is more beneficial than the policy against the deceiver's true type---deception will occur. We measure the various metrics of deception during the interaction between the DoM$(1)$ sender and DoM$(0)$ receiver:  false beliefs (Eq.~\ref{eq:false_beliefs}) and deviation from expected reward (Eq.~\ref{eq:cumulative_reward_difference}). The values of these are shown in Fig.~\ref{fig:mixed_motive_deception_illustration}(B,C). First, the actions of the DoM$(1)$ sender causes the DoM$(0)$ receiver to form false beliefs,as illustrated in Fig.~\ref{fig:mixed_motive_deception_illustration}(B). In this case the receiver's beliefs are misplaced, as: $P(\psi_\mu = \text{random}) > P(\psi_\mu \in \{0.1,0.5\})$, throughout the interaction ($\bm{t_D} = T$). The expected reward from the DoM$(0)$ receiver is computed via Eq.~\ref{eq:expected_reward_for_given_belief} and presented in Fig.~\ref{fig:mixed_motive_deception_illustration}(C) as the striped bars. These can then be compared to the observed reward (lower, non striped bars). This figure shows that, on average, the DoM$(1)$ sender retains between 40\% ($\psi = 0.1$) to 70\% ($\psi = 0.5$) more reward than the receiver expects without being detected or causing the DoM$(0)$ to change its behaviour.

The deceptive behaviour of the DoM$(1)$ sender in the IUG task illustrates the main steps of deception with ToM,
a pattern that is not unique to the IUG game, but has also been reported in other work, for example, \citet{alon2023dis}. The deceiver is playing a gambit, incurring some cost to itself, at the expense of installing false beliefs which it later exploits. The victim's inability to reason about this complex behaviour causes it to act according to the the deceiver's plan.

\section{Overcoming Deception with Limited Computational Resources}

The central idea of this paper is that despite limitations in opponent modelling, the victim can resist and detect deception. This is realised by assessing the (mis-)match between the \emph{expected} (based on the victim's lower DoM) and \emph{observed} behaviour of the opponent---a form of heuristic behaviour verification, or prediction error. For example, consider a parasite, masquerading as an ant to infiltrate an ant colony and steal food. While its appearance may mislead the guardian ants, its behaviour---feasting instead of working---should trigger an alarm. Thus, even if the victim lacks the cognitive power to characterise or understand the manipulation, they can detect the strategy through a form of conduct violation. If the victim detects such a discrepancy, they can conclude that the observed behaviour is generated by an opponent that lies outside their world model. In principle, there is a wealth of possibilities for mismatches, including incorrect prior distributions over parameters governing the DoM$(-1)$ behaviour or the utilities. Here we assume that the only source of poor model capacity is the limited DoM level. For a level $k$ agent, we denote the set of modelled levels as $\Theta_k  \subset \Theta$, and the set of unmodelled levels as $\Theta_\aleph$. For a DoM$(k)$ agent, all the possible DoM$(k-1)$ models are in $\Theta_k$. Thus, a mismatch between the observed and expected behaviour implies that the unknown agent has DoM level \emph{different} from $(k-1)$. This is a pivotal concept, allowing the agent to engage with a known, yet unmodeled, opponent. Once the behavioural mismatch has been identified, the victim has to decide how to act. Defensive counter-deceptive behaviour should consider the observation that an out-of-model deceiver can predict the victim. Thus, the victim's policy could be aimed at hurting the deceiver (at a cost to the victim), to deter them from doing so.

\subsection{Detecting and Responding to Deception with a Limited Opponent Model}

Detecting abnormal, and potentially risky, behaviour from observed data is related to Intrusion Detection Systems (IDS). This domain assumes that ``behaviour is not something that can be easily stole[n]'' \citep{salem_survey_2008}. Thus, any atypical behaviour is flagged as a potential intruder, alarming the system that the observed user poses a risk. Several methods have been suggested to combat a masquerading hacker \citep{salem_survey_2008}. Inspired by these methods, we augment the victim's inference with an $\aleph$-mechanism. This mechanism, $f(\Theta_k, h^t)$, evaluates the opponent's behaviour against the expected behaviour, based on the agent's DoM level and the history. The expected behaviour includes the presumed opponent's response to the agent's actions, similar to the simulated policy in Equation \ref{eq:simulated_opponent_behaviour}. The $\aleph$-mechanism returns a binary vector of size $|\Theta_k|$ as an output. Each entry represents the $\aleph$-mechanism's evaluation per opponent type: $\theta_\nu \in \Theta_k$. The evaluation either \textit{affirms} or \textit{denies} that the observed behaviour (discussed next) sufficiently matches the expected behaviour of each agent type. A critical issue is that, as in \citet{liebald2007proactive}, we allow the deceiver to be aware of the detailed workings of the $\aleph$-mechanism, and so be able to avoid detection by exploiting its regularities. This renders many methods impotent for detecting deception. Nonetheless, to gain an advantage from its deception, the deceiver will have to deviate from typical behaviour at some point (otherwise it will just emit the same actions against which the victim is best responding). For example, in the zero-sum game, the deceiver's behaviour shifts from typical behaviour for a given opponent policy to non-typical behaviour for the same behavioural type. This should alert the victim that they might be engaging with an unmodelled agent. Of more consequence, at least in a non-fully cooperative game, is that the victim will get less cumulative reward than it expects based on Equation \ref{eq:expected_reward_for_given_belief}. Thus, if the victim's \emph{actual} reward deviates (statistically) from the expected reward, they can estimate that the opponent is not in their world model.

\subsubsection{$\aleph$-Mechanism}
The behaviour verification mechanism assesses whether the observed opponent belongs to the set of potential opponents using two main concepts: typical behaviour monitoring and the expected reward. The concept of typical behaviour arises from Information Theory (IT), where the \emph{typical set} is defined as the set of realisations from a generative model with empirical and theoretical frequencies that are sufficiently close according to an appropriate measure of proximity. This measure is often used to compute asymptotic values such as channel capacity. In this work we use it to decide if a finite sequence of observations is a typical behaviour given an opponent model. If an observed trajectory of behaviour does not belong to an expected typical set there is a high probability that the behaviour was not generated by the expected model. The second component confirms that the observed reward is consistent with the expected reward for each partner type. Strong typicality is often used to compute the asymptotic properties of a communication channel. Each type acts in a certain way, depending on the duration of the interaction and the responses of the primary agent. Thus, the inferring agent can compute a distribution over the responses of each type in its opponent set (Eq.~\ref{eq:simulated_opponent_behaviour}). We present two different approaches to compute this size, depending on the nature of the modelled opponent. We begin with presenting a $\delta$-strong typicality model. Formally, for each time $t \in [0,T]$, let $\hat F^t_{h^t}(a'_\nu)$ denote the empirical likelihood of an opponent action in the corresponding history set $h^t$, defined as $\hat F^t_{h^t}(a'_\nu) = \sfrac{N(a'_\nu)}{|h^t|}$, where $N(a'_\nu)$ is the number of times the action $a'_\nu$ appears in the history set. Let $F^t_{\theta_\nu}(a'_\nu)$ be the theoretical probability that an agent with type $\theta_\nu$ will act $a'_\nu$ at time $t$ given history $h^{t-1}$. A \emph{$\delta$-strongly typical set} of a trajectory, for agent with type $\theta_\nu$, is the set:
\begin{equation}
Y^t_\delta(\theta_\nu) = \{h^t: |\hat F^t_{h^t}(a'_\nu) - F^t_{\theta_\nu}(a'_\nu)| \leq \delta \cdot F^t_{\theta_\nu}(a'_\nu)\}
\label{eq:typical_set}
\end{equation}
The parameter $\delta$ governs the size of the set, which in turn affects the sensitivity of the mechanism. It can be tuned using the nested opponent models to reduce false positives. We analyse the sensitivity of the mechanism to various values of $\delta$ in the next section. However, since the deceiving agent model is absent from this ``training'' set, this parameter cannot be tuned to balance true negatives. An additional issue with setting this parameter is its lack of sensitivity to history length; the distance between the expected behaviour according to history (theoretical) and the empirical (the actual observed behaviour) reduces with $t$, but is quite high when $t$ is small. We address this issue by making it trial-dependent, denoted by $\delta(t)$. We discuss task-specific details in Appendix \ref{app:zero_sum_description}. However, this condition is only useful when the observations are independent and identically distributed (iid) variables. This holds for a random sender in the IUG task, or the agents depicted in the Row/Column game (Appendix \ref{app:zero_sum_description}). In many multi-agent interactions, the actions at time $t$ depend on the history up to that point. For example, as illustrated in Fig.~\ref{fig:iug_strategic_behaviour_illustration}, the DoM$(-1)$ threshold sender behaviour (policy) changes each trial. To overcome this issue we present a second computational method to estimate a form of sequential typicality. Inspired by the work of \citep{wang_anomaly_2012}, we propose a gzip-based  \citep{deutsch1996gzip} algorithm to compute the sequential typicality set. This is a sampling-based algorithm, depicted in Alg.~\ref{alg:gzip_compression}. The core idea is to generate independently sampled trajectories for each simulated opponent policy (Eq. \ref{eq:simulated_opponent_behaviour}), adding  $\hat o_{\nu,n}^t \sim \hat\pi^t_{\nu_{k-1}}(\hat \theta^t_{\nu_{k-1}})$ at trial $t$ to a previously sampled trajectory $\hat o_{\nu,n}^1, \hat o_{\nu,n}^2, \dots, \hat o_{\nu,n}^{t-1}$. These sampled trajectories are collected throughout the interaction, such that the ``trial'' set is 
\begin{equation}
    \hat{D^{0:t}_\nu} = \{[\hat o_{\nu,n}^0, \hat o_{\nu,n}^1, \dots, \hat o_{\nu,n}^t]| \hat o_\nu^\tau \sim \hat\pi^\tau_{\nu_{k-1}}(\hat \theta^\tau_{\nu_{k-1}}), n \in [N]\}
    \label{eq:trajectories_set}
\end{equation}
Each element of this set of sampled sequential trajectories is then separately compressed using the gzip algorithm, generating a set of compression ratios $C^t = \{c^t_1, c^t_2, \dots c^t_N\}$, where $N$ is the sample size and $c^t_n$ is the compression ratio of the $n$th sampled trajectory. The observed sequence is also compressed, and its compression ratio $c^t_O$ is compared to the distribution of the sampled trajectories compression ratio set. The parameter $\delta$ is used to set the percentile at which the comparison takes place. The observed trajectory is classified as typical if $c^t_O$ is larger or equal than the $\delta$ percentile of $C^t$ or smaller or equal to the $1-\delta$ percentile of $C^t$. In this context $\delta$ can be seen as a surrogate p-value, governing the location of the observed sequence compared to the theoretical ones. Given that the set $C^t$ can be thought of as a sample of compression ratios, the width (uncertainty) of this surrogate ``confidence-interval'' is governed by the size of the sample size $N$. Small $N$ may lead to high uncertainty which damages the ability of this component to properly asses if the observed compression ratio fits in the theoretic set (error reduces proportionally to $\sqrt{N}$). In this work we used $N=200$ meaning that the error was reduced by a factor of more than $10$. The typicality-based component of the $\aleph$-mechanism, denoted by $Z^1(\Theta_k, h^t, \delta)$, outputs a binary vector, per trial $t$. Each entry in the vector indicates whether or not the observed sequence belongs to the typical set of $\theta_\nu \in \Theta_k$. The algorithm is defined by the environment (opponent models).

The second component, denoted by $Z^2(\Theta_k, h^t, r^t, \omega)$, verifies the opponent type by comparing the expected cumulative reward to the observed cumulative reward. In any MARL task, agents are motivated to maximise their utility. In mixed-motive and zero-sum games, the deceiving agent increases its portion of the joint reward by reducing the victim's reward. This behaviour will contradict the victim's expectations to earn more (due to the assumption that it has a higher DoM level). Since gaining (subjective) utility is the victim's ultimate aim, we can expect the victim to be particularly sensitive to behaviour that leads to deviation from the expected reward. A similar idea was proposed by \citet{oey2023designing}, where the authors suggested the deception detection mechanism in humans is governed by the possible payoffs. Due to the coupling between the agent's reward and its actions, this component is based on the history-conditioned expected reward $\hat r_\mu^t$, by averaging the expected actions and reactions per behavioural type. Formally, the expected reward, per opponent's behavioural type, is:
\begin{equation}
    \hat r_\mu^t(\theta_{\nu_{k-1}}) = E_{a_\mu^t \sim \pi^t_{\mu_k}}[E_{a_\nu^t \sim \hat\pi^t_{\nu_{k-1}}}(u_\mu(a_\mu^t,a_\nu^t))|h^{t-1}]
    \label{eq:expected_reward_per_type}
\end{equation}

Similarly to the gzip typicality component, this component also utilises counterfactuals to asses whether or not the observed sequence of rewards ``fits'' a presumed generative model. At each trial, the component samples actions from the simulated policy of each modelled type (Eq. \ref{eq:simulated_opponent_behaviour}) and, using the victim's policy, computes a vector of sampled rewards. These rewards are appended to previously sampled rewards, yielding the expected reward set:
\begin{equation}
    \hat{R^{0:t}_\nu} = \{[\hat r_{\mu,n}^0, \hat r_{\mu,n}^1, \dots, \hat r_{\mu,n}^t]|a_\mu^t \sim \pi^t_{\mu_k}, {a_\nu^t \sim \hat\pi^t_{\nu_{k-1}}}, n \in [N]\}
    \label{eq:sampled_reward_set}
\end{equation}
At each trial, this component verifies if the cumulative \emph{observed} reward: $\sum_{i=1}^t r^i_\mu$ belongs to the set of \emph{expected} cumulative rewards:
\begin{equation}
\hat{CR^{t}_\mu} = \{\sum_{i=1}^t \hat r_{\mu,n}^i| n\in [N]\}
\label{eq:set_of_sampled_cumulative_rewards}
\end{equation}
This is determined by a parameter $\omega \in [0,1]$ that is used similarly to $\delta$ in the gzip components, determining the percentiles that the observed cumulative reward needs to satisfy to be counted as `typical' reward. This component's output is similar to the output of $Z^1(\Theta_k, h^t)$, namely a vector of size $|\Theta_k|$ with each entry is a binary variable, indicating if the observed reward is within acceptable bounds, implying whether the reward could have been generated by a $\theta_\nu \in \Theta_k$ opponent. Notably, if the expected reward is higher than the upper limit, this component is also activated, even though such an event benefits the victim. Of course, this component may be tuned to alert only when the expected reward is too low. The two components are correlated, as $\mu$'s reward, monitored by $Z^2$ is a function of $\nu$'s actions, which are monitored by $Z^1$. We combine the output of the components using element-wise logical conjunction recursively: $f^t = f^{t-1} \wedge (Z^1(\cdot) \wedge Z^2(\cdot))$. The recursive update, similar to the Bayesian update, is inspired by the assumption that agents cannot change their type during an interaction \citep{gmytrasiewicz_framework_2005}---meaning that if a certain type is excluded by the $\aleph$-mechanism at time $t$ it cannot be verified in the future. The full mechanism is described in Algorithm \ref{alg:aleph_mechanism}.
The output, a binary vector, is then one of the inputs to the $\aleph$-policy.
\begin{algorithm}
\caption{$\aleph$-mechanism}
\label{alg:aleph_mechanism}
\hspace*{\algorithmicindent} \textbf{Input}: $f^{t-1}, \Theta_k, h^{t}, r^t, \delta(t), \omega$ \\
\hspace*{\algorithmicindent} \textbf{Output}: $f^t$ 
\begin{algorithmic}[1]
\Procedure{$\aleph$-Mechanism}{$f^{t-1}, \Theta_k, h^{t}, r^{t}, \delta(t), \omega, N$}
\State $x \gets Z^1(\Theta_k,h^t, \delta(t), N)$ \Comment{Environment dependent}
\State $y \gets Z^2(\Theta_k,h^t, r^t, \omega)$
\State $f^t \gets f^{t-1} \wedge (x \wedge y)$
\State \textbf{return} $f^t$
\EndProcedure
\Procedure{$Z^1_{\text{gzip}}$}{$\Theta_k,h^t, \delta(t), N$} \Comment{From Alg. \ref{alg:gzip_compression}}
    \State Compute $Y^t_{\delta(t)}(\theta)$  \Comment{From \ref{eq:typical_set}}
    \State $x\gets h^t \in Y^t_{\delta(t)}(\theta)$ 
    \State \textbf{return} $x$ 
\EndProcedure

\Procedure{$Z^2$}{$\Theta_k, h^t,r^t, \omega$}
    \State Sample counterfactual reward set \Comment{From \ref{eq:sampled_reward_set}}
    \State Compute cumulative reward set \Comment{From \ref{eq:set_of_sampled_cumulative_rewards}}
    \State Compute cumulative empirical reward: $CR^t_\mu = \sum_{i=1}^t r^i_\mu$
    \State $q_\omega, q_{1-\omega} \gets$ Compute $\omega$ and $1-\omega$ percentile of $\hat{CR^{t}_\mu}$
    \State $y \gets q_\omega \leq CR^t_\mu \leq q_{1-\omega}$
    \State \textbf{return} $y$
\EndProcedure
\end{algorithmic}
\end{algorithm}

\begin{algorithm}
\caption{gzip-compression}
\label{alg:gzip_compression}
\hspace*{\algorithmicindent} \textbf{Input}: $\Theta_k,h^t, \delta,  N$ \\
\hspace*{\algorithmicindent} \textbf{Output}: $x$ 
\begin{algorithmic}[1]
\Procedure{$Z^1_{\text{gzip}}$}{$\Theta_k,h^t, \delta, N$}
\State $c^t_O \gets$ Compute compression ratio of observation from $h^t$ using gzip
\For{$\theta^t_{\nu_{k-1}} \in \Theta_k$}:
\State Generate sample $D^t$ of size $N$ observations $o_i^t$ from $\hat\pi^t_{\nu_{k-1}}(\hat \theta^t_{\nu_{k-1}})$
\State Append $D^t$ to $D^{0:t-1}$
\EndFor
\State $C^t \gets$ Compute compression ratio for each sampled trajectory $\hat o_{\nu,n}^t \in D^{o:t}$ using gzip
\State $q_\delta, q_{1-\delta} \gets$ Compute $\delta$ and $1-\delta$ percentile of $C^t$
\State $x \gets q_\delta \leq c^t_O \leq q_{1-\delta}$
\State \textbf{return} $x$
\EndProcedure
\end{algorithmic}
\end{algorithm}

\begin{figure}
    \centering
    \includegraphics[scale=.10]{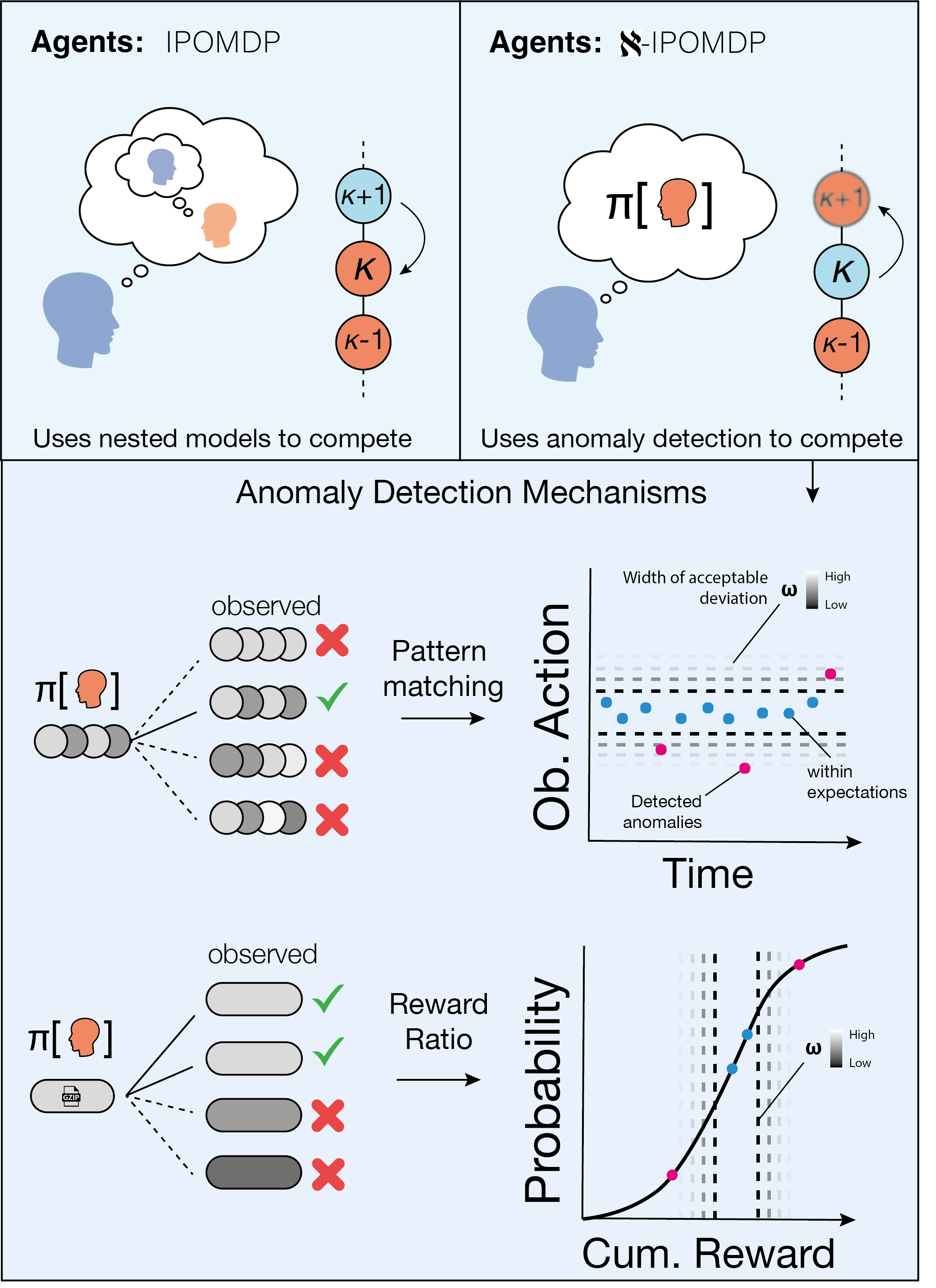}
    \caption{\textbf{Illustration of the $\aleph$-mechanism algorithm} (Top Left) In IPOMDP, DoM$(k)$ agents use their nested model to compute best-response against DoM$(k-1)$ agents, taking advantage of their superior mentalising abilities. (Top Right) $\aleph$-IPOMDP augmentations allow agents to detect when their partners might be hierarchically superior, utilizing anomaly detection methods to avoid exploitation. (Bottom) The $\aleph$-mechanism combines expectation-observation monitoring with  typicality (policy predictions) to verify that the observed agent acts ``as expected''. This can vary by agents, with individual differences dictating how much `evidence' of anomalies is required before the mechanism is triggered, and also how narrow or broad the typical set is.}
    \label{fig:enter-label}
\end{figure}

\subsubsection{$\aleph$-Policy}
The agent's behaviour is governed by its $\aleph$-policy (Algorithm \ref{alg:aleph_policy}). This takes as its input the output of the $\aleph$-mechanism and the beliefs about the opponent's type and generates an action. If the opponent's behaviour passes the $\aleph$-mechanism for at least one type, the agent uses its DoM$(k)$ policy, in this work a SoftMax policy. Here we use the IPOMCP algorithm for the Q-value computation \citep{hula_monte_2015}, but any planning algorithm is applicable. In the case that the opponent's behaviour triggers the $\aleph$-mechanism, the victim's optimal policy switches to an out-of-belief (OOB) policy. While lacking the capacity to simulate the external opponent, the $\aleph$-policy utilises the property highlighted above: the victim knows that the deceiver is of an unknown DoM level. If the deceiver's DoM level is higher than the victim's, it means that the deceiver can fully simulate the victim's behaviour. Hence, if the OOB policy derails the opponent's utility maximisation plans, the opponent will avoid it. Building on our axiomatic framework, we define an effective OOB policy as one that neutralizes the deceiving agent's utility gain (Axiom 1), thereby disincentivizing it from engaging in deceptive behaviour. The best response inevitably depends on the nature of the task. We illustrate generic OOB policies for two different payout structures: zero-sum and mixed-motive. In zero-sum games, the Minimax algorithm \citep{shannon1993programming} computes the best response in the presence of an unknown opponent. This principle assumes that the unknown opponent will try to act in the most harmful manner, and the agent should be defensive to avoid exploitation. While this policy is beneficial in zero-sum games it is not rational in mixed-motive games; it prevents the agent from taking advantage of the mutual dependency of the reward structure. In repeated mixed-motive games, several policy prescriptions have been suggested as ways of deterring a deceptive opponent. An agent following the Grim trigger policy \citep{friedman1971non} responds to any deviation from cooperative behaviour with endless anti-cooperative behaviour, even at the risk of self-harm. While being efficient at deterring the opponent from defecting, this policy has its pitfalls. First, it might be that the opponent's defective behaviour is by accident and random (i.e., due to SoftMax policy), and so endless retaliation is misplaced. Second, if both players can communicate, then a warning shot is a sufficient signal, allowing the opponent to change their deceptive behaviour. This requires a different model, for example, the Communicative-IPOMDP \citep{gmytrasiewicz2019optimal}. However, in this work, we illustrate how the possibility of a Grim trigger policy suffices to deter a savvy opponent from engaging in deceptive behaviour.
\begin{algorithm}
\caption{$\aleph$-policy}
\label{alg:aleph_policy}
\hspace*{\algorithmicindent} \textbf{Input}: $b^t_{\mu_k}(\theta_{\nu_{k-1}}), f$ \\
\hspace*{\algorithmicindent} \textbf{Output}: $a^{t+1}$ 
    \begin{algorithmic}[1]
        \Procedure{$\aleph$-Policy}{$b^t_{\mu_k}(\theta_{\nu_{k-1}}), f$}
            \If {$f \neq 0$}:            
            \State $a^{t+1} \sim \pi^t_{\mu_k}(b^t_{\mu_k}(\theta_{\nu_{k-1}}))$
        \Else \ 
            \State $a^{t+1}$ is sampled from OOB policy
        \EndIf
  
        \State \textbf{return} $a^{t+1}$ 
        \EndProcedure    
    \end{algorithmic}
\end{algorithm}

\subsubsection{Mixed-Motive Game}
\begin{figure}[htbp]
    \centering
    \includegraphics[scale=.067]{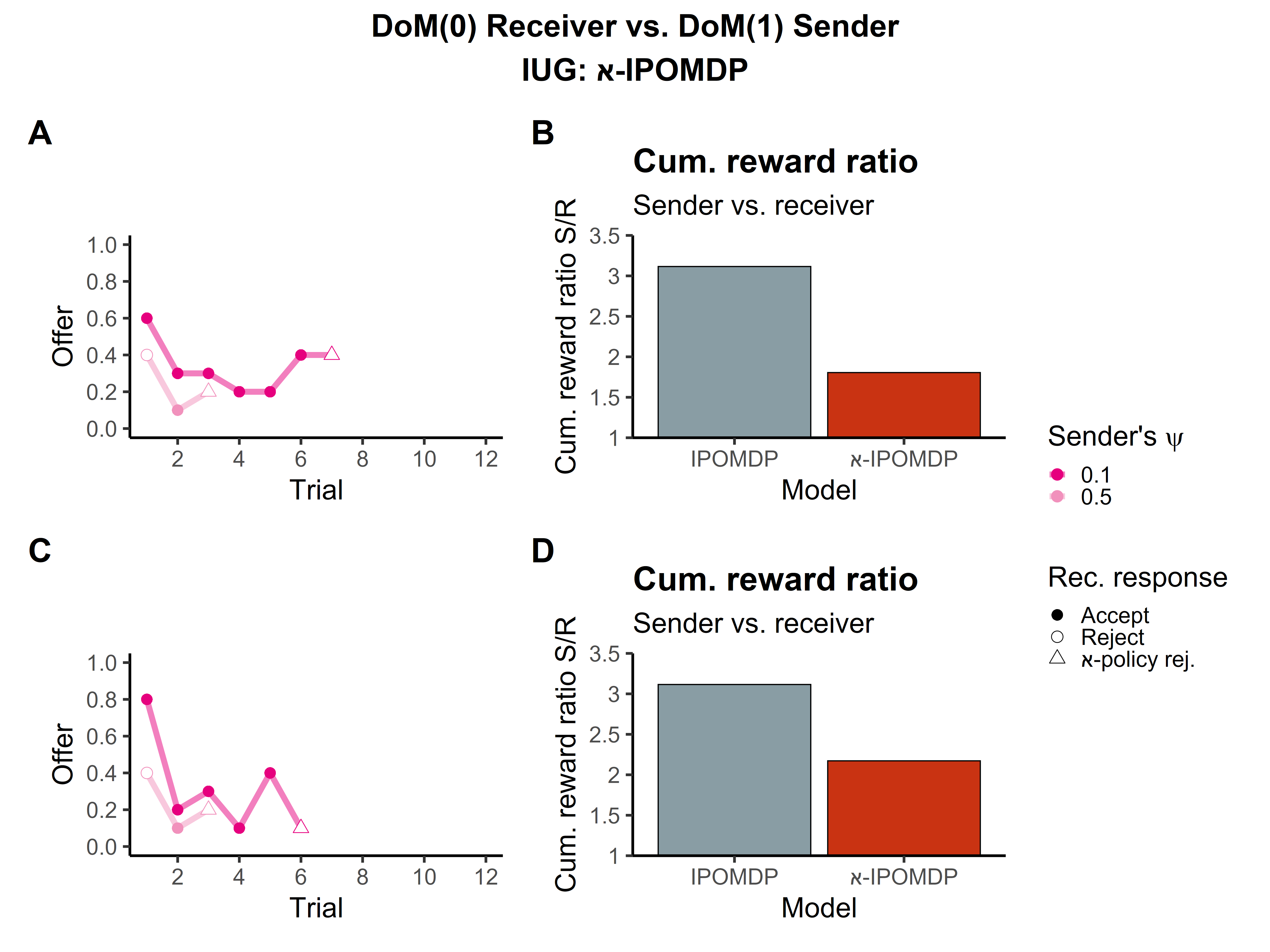}    
    \caption{\textbf{Mitigation of deception in IUG with $\aleph$-IPOMDP}: 
    Points represent offers from the sender to the receiver across trials, biased by the sender's threshold. Points are shaded white if the receiver rejects the offer, triangular points indicate that the rejection is caused by the $\aleph$-mechanism, effectively terminating the interaction. Lines and points are visible while the $\aleph$-mechanism is off. \\
    \textbf{Top row  $\delta=0.1$, $\omega=0.3$} --- \textbf{(A)}  Notably, both DoM$(1)$ senders masquerade as being random. However, their ability to execute the ``random'' behaviour ruse is limited by both $\aleph$-mechanism components. First, the cumulative reward has to satisfy the off-policy counterfactual reward component. Next, the variability of the offers is higher than in the IPOMDP case, respecting the typicality component. Ultimately, the deceiver's policy triggers the average reward monitoring ($Z^2$) component, as the observed reward is lower then expected (marked by the truncation of the line and points). \textbf{(B)} Cumulative reward ratio for the sender vs.~the receiver. The $\aleph$-IPOMDP reduces the cumulative reward ratio (sender/receiver) by more than $40\%$. \\
    \textbf{Bottom row $\delta=0.3$, $\omega=0.3$} --- \textbf{(C)} When it is constrained by narrower strong typicality set $(Z^1)$ bounds, the DoM$(1)$ with low threshold terminates the interaction faster than before, triggering this component after 6 trials, while the high threshold sender acts similarly.\textbf{(D)} Even when the interaction is shorter, the reward ratio is still reduced compared to the case of the conventional IPOMDP.}    
    \label{fig:iug_aleph_illustration}
\end{figure}

Repeating the simulation with the $\aleph$-IPOMDP framework shows how a power imbalance is diluted via the detection and retaliation of the $\aleph$-mechanism, and via the Grim Trigger-based $\aleph$-policy. The effect of the $\aleph$-IPOMDP is illustrated in Figure~\ref{fig:iug_aleph_illustration} depicting how the deceiver's behaviour is affected by the  $\aleph$-mechanism and the $\aleph$-policy (compare Figures \ref{fig:mixed_motive_deception_illustration}(A)~and~\ref{fig:iug_aleph_illustration}(A)). Each component of the $\aleph$-mechanism limits the freedom of action of the deceiver if they are to avoid the consequences of detection. As illustrated in Figure \ref{fig:iug_aleph_illustration}, these constraints have different effects on the two deceptive DoM(1) senders in the IUG task:
while both DoM(1) senders attempt to masquerade as random senders, their behaviour drastically changes compared to the IPOMDP case. First, since the expected cumulative reward component constrains them, these senders can no longer make only low offers; their offers vary to ensure that the cumulative reward is within the bounds dictated by $\omega$. Second, the typicality monitoring ($\delta$) forces them to diversify their offers to avoid over-efficient compression. These adapted behaviours are presented for two sets of $\delta,\omega$ in Fig.~\ref{fig:iug_aleph_illustration}(A,C)

Overall, our proposed mechanism reduces the income gap between the agents, as illustrated in Fig.~\ref{fig:iug_aleph_illustration}(B,D). Thus it limits the ability of a deceptive, higher DoM agent to take advantage of a limited computational victim.

\subsubsection{Effect of $\aleph$-Mechanism Parameters}
\label{sec:aleph_grid_parameters}
\begin{figure}[htbp]
    \centering
    \includegraphics[scale=.086]{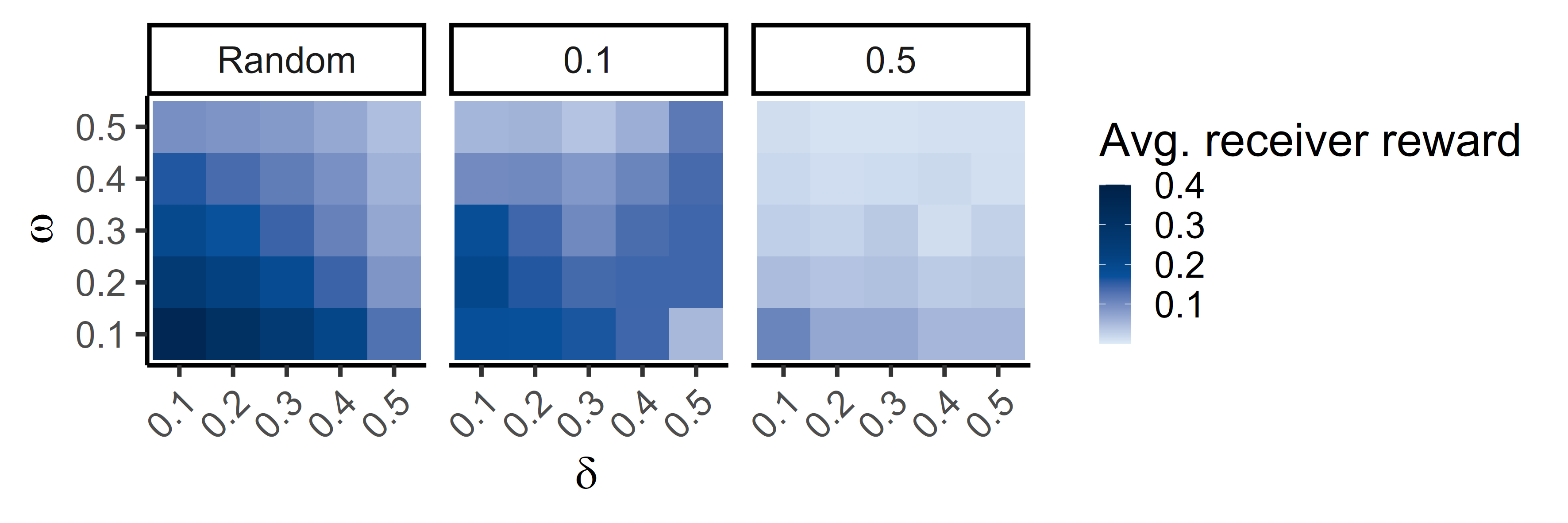}    
    \caption{\textbf{Effect of $\aleph$-mechanism parameters on IUG outcome for DoM$(0)$ receiver interacting with random and DoM$(1)$ sender}: Measuring the receiver's average reward (averaged across 20 simulations) in the IUG with different, fixed, values of $\omega$ (y-axis) and $\delta$ (x-axis) reveals how the parameters of the $\aleph$-mechanism affect the interaction. When engaged with a random sender (Left), tight constraints lead to an overactive $\aleph$-mechanism that nullifies the interaction. This is because the random sender does not alter its behaviour in response to rejections by the receiver. In this case, the DoM$(0)$ receiver needs to be very flexible in its $\aleph$-mechanism to collect a reward. As evident the tighter the bounds (higher parameters) the more likely it is that the mechanism with mistakenly flag the sender as non-random, terminating the interaction. This plot show that the $\aleph$-mechanism is more likely to be activated by the reward monitoring. However, while high values of $\omega$ (narrow reward bounds) suppress the $Z^1$ component, when these bounds are quite wide, the strong typicality component is more likely to activated ($\omega \in [0.1, 0.2]$). Moreover, the $Z^1$ mechanism is more likely to be activated when interacting with the higher threshold sender}
    \label{fig:iug_aleph_grid_results_receiver_reward}
\end{figure}

The size of the inequality reduction is a function of the $\aleph$-mechanism parameters. Narrowing the expected reward bounds, using a small $\omega$,  forces the deceiver to make offers that are closer to the offers that the agent it is masquerading as would make, reducing the size of the set of available actions (to avoid alerting the victim). Setting larger values of $\delta$ limits the deceiver's ability to repeat the same offer several times. The combination of the two determines the outcome of the game. However, this rigidity may harm the victim's performance when interacting with a genuinely random agent, as could in fact be the case in the IUG task. Hence, setting these parameters requires a delicate balance between false and true negatives. To examine this, we simulated the DoM$(0)$ $\aleph$-IPOMDP receiver against a DoM$(1)$ sender and a random sender over a grid of $5$ different values of $\delta$ and $5$ values of $\omega$. Each design was simulated 50 times with different random seeds. We first discuss the effect of the parameters on the receiver's reward, as the goal of the $\aleph$-IPOMDP is to improve the victim's performance via deterrence. The average reward of the receiver is depicted in Fig.~\ref{fig:iug_aleph_grid_results_receiver_reward}. We begin with the results against a random sender (left-most panel)---here we clearly see the detrimental results of high levels of $\omega$ and $\delta$---aimed at verifying ``genuine'' random behaviour. As shown in Fig.~\ref{fig:iug_aleph_grid_results_receiver_reward} (left column), there is a partial trade-off between $\delta$ and $\omega$ when interacting with a random sender. The higher these parameters, the more sensitive the $\aleph$-mechanism is to deviations from ``truly'' random behaviour. In turn, this causes the $\aleph$-mechanism to flag a genuine random sender as being non-random, thus activating the $\aleph$-policy and reducing the receiver's average reward per trial compared to the ``n\"aive'' case. 

The effect of the parameters is also substantial when engaged with the DoM$(1)$ sender. Here we compare the average reward of the $\aleph$-IPOMDP receiver to the ``sucker payoff'' ($0.1$). While the framework as a whole limits the DoM$(1)$ sender's ability to deceive the DoM$(0)$ receiver, certain settings allow the DoM$(1)$ larger wiggle room. As alluded to above, while the overall effect of the $\aleph$-mechanism is a reduction in the reward ratio, the causal effect differs between the threshold senders.
The high DoM$(1)$ threshold sender ($\psi=0.5$) only engages with the DoM$(0)$ $\aleph$-agent when the parameters are low, as any other setting coerces it to make offers that are detrimental to it. This avoidance of engagement illustrates how the $\aleph$-IPOMDP framework serves as a deterrence mechanism---the sender is aware of the ``commitment'' of the receiver to ``cut off its nose to spite both faces'', i.e., the mechanism serves as a credible threat, dissuading the high threshold sender from engaging with it.

On the other hand, in order to maximize reward, the low threshold ($\psi=0.1$) DoM$(1)$ sender alters its behaviour (compared to the deceptive baseline) to gain reward, where possible. It continues to masquerade  as being random, but its policy adheres to the statistical constraints imposed by the $\aleph$-mechanism, until it is compelled to make offers that do not benefit it, in which case its policy triggers the $\aleph$-mechanism, ending the interaction. This effect is evident in Fig. \ref{fig:iug_aleph_grid_results_receiver_reward} (middle column). Some combinations of  $(\omega,\delta)$ do not harm the receiver's outcome substantially compared to the non-regulated game, but others reduce the receiver's reward by more than $50\%$. This phenomenon illustrates Goodhart's law---the deceiver uses its nested model to estimate, using simulation, how to avoid detection, at a cost. While the $\aleph$-IPOMDP framework fails to coerce the deceiver into  disclosing its true type fully (and in fact it encourages it to learn a new deceptive move) it succeeds in mitigating the effect of deception, namely decreasing the gain from deceptive behaviour and reducing the reward gap.

\begin{figure}
    \centering
    \includegraphics[width=0.6\linewidth]{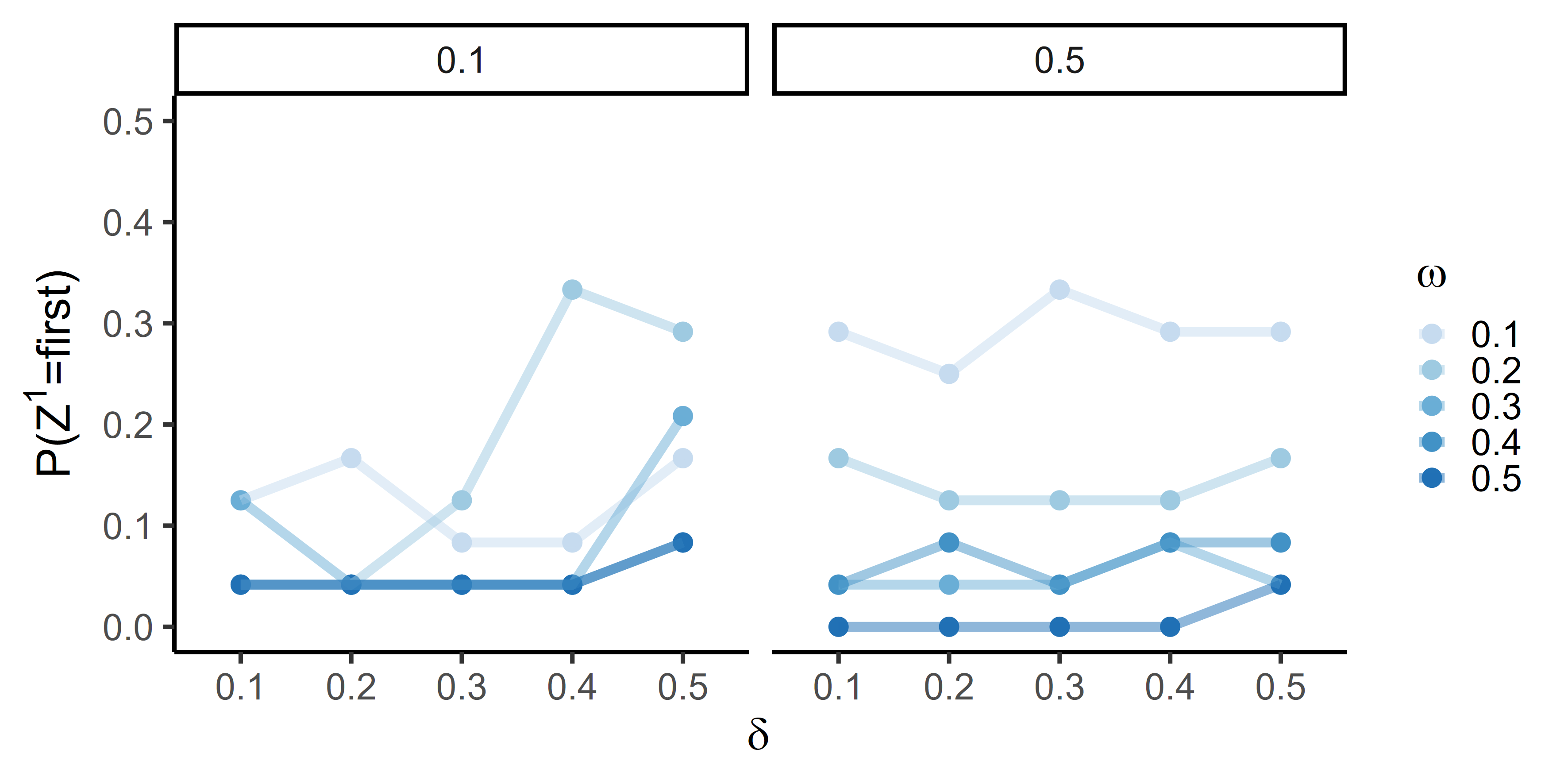}
    \caption{\textbf{Probability of activation by $Z^1$}. The points show the probability that the $\aleph$-Mechanism is activated by the strong-typicality component prior to being activated by the reward monitoring component $(Z^2)$. Lines indicate the size of the $\omega$ and each column represents a different sender threshold. This plot show that the $\aleph$-mechanism is more likely to be activated by the reward monitoring. However, while high values of $\omega$ (narrow reward bounds) suppress the $Z^1$ component, when these bounds are quite wide, the strong typicality component is more likely to activated ($\omega \in [0.1, 0.2]$). Moreover, the $Z^1$ mechanism is more likely to be activated when interacting with the higher threshold sender}
    \label{fig:iug_aleph_mechanism_activation_by_component_and_threshold}
\end{figure}

\paragraph{$\aleph$-Mechanism Components Effect}
\label{sec:aleph_mechanism_components_effect}
As illustrated in Fig.~\ref{fig:iug_aleph_illustration}, the $\aleph$-mechanism disturbs the deceptive behaviour of the DoM$(1)$ sender, causing it to act in a more random manner, avoiding detection as long as possible. This modified deceptive behaviour is still picked by the $\aleph$-mechanism, but which of the two components is more likely to detect the mismatch between the expected and observed behaviour is triggered first? We answer this question by plotting the probability that the $\aleph$-mechanism is triggered first by the strong-typicality component ($Z^1$) by sender threshold, depicted in Fig.~\ref{fig:iug_aleph_mechanism_activation_by_component_and_threshold}. In this figure we depict the probability that the $\aleph$-mechanism is activated by $Z^1$ before being triggered by $Z^2$. This plot shows that overall the mechanism is more susceptible to deviation from expected reward rather than to deviation from expected compression ratio, but this effect differs by the sender's threshold and associated behaviour. We included an analysis by trial in Appendix~\ref{fig:iug_aleph_mechanism_activation_by_component_threshold_and_trial}.

\paragraph{$\aleph$-Mechanism False Positives}
\label{sec:false_positives}

\begin{figure}[h!]
    \centering
    \includegraphics[width=0.8\linewidth]{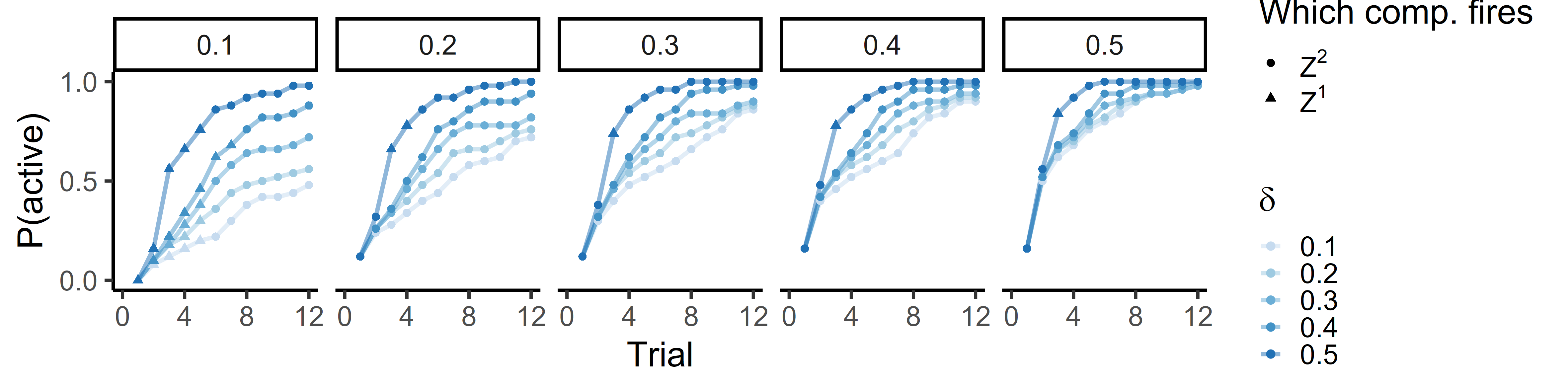}
    \caption{\textbf{$\aleph$-mechanism false positive probability}. The $\aleph$-mechanism misdetects a random sender as non-random, causing the $\aleph$-IPOMDP agent to wrongly terminate the interaction. This plot depicts this false-positive probability (y-axis) as a function of the trial number (x-axis) and the $\aleph$-mechanism components parameters---$\omega$ (columns) and $\delta$ (colours). When the parameters are low enough, the probability of false-positive is lower than $0.5$ (for the case of $\omega=0.1, \delta=0.1$), but spikes up to almost $1$ when the parameters are high. This false identification is caused by an overly sensitive and narrow window of acceptable deviation from expected outcomes.}
    \label{fig:aleph_mechanism_false_positive}
\end{figure}

While effective in deterring higher DoM agents from engaging in protracted deceptive behaviour, the $\aleph$-mechanism may backfire when interacting with a genuine agent---in this case, the random sender \citep{alon2023between}. This false detection depends on the parameters of both components. We begin with an overall analysis of false detection illustrated in Fig.~\ref{fig:aleph_mechanism_false_positive}. This plot depicts the probability that the mechanism will falsely activate when engaging with a random sender as a function of the interaction duration. As evident the tighter the bounds (higher parameters) the more likely it is that the mechanism with mistakenly flag the sender as non-random, terminating the interaction. This plot shows evidence that the $\aleph$-mechanism is overly aggressive. Future work may resolve this issue by balancing false positives and negatives. This finding better explains the poor results depicted in Fig.~\ref{fig:iug_aleph_grid_results_receiver_reward}(Left panel). Crucially, as in the case of the truly non-random sender the mechanism is more sensitive to deviations from strong typicality in early trials and is more vulnerable to misclassifications by the average reward monitoring in later trials.

Thus, in the spirit of a ``no free lunch'' theorem, our results show that the best set of parameters against the low threshold DoM$(1)$ sender is not best against the high threshold  DoM$(1)$ sender nor against a random one. However, on average, certain sets of parameters yield the desired outcome---mitigating deceptive behaviour.

\section{Discussion}

We imbued agents with the ability to assess whether they are being deceived, without (at least fully) having to conceptualise how. To achieve this we augmented Bayesian inference within the $\aleph$-IPOMDP mechanism with environmental and anomaly detection mechanisms. The $\aleph$-policy of these agents is the ability to infer that they are facing an agent outside their world model that threatens to harm them. The net result is a more equitable outcome. This framework offers a new conceivable algorithm to tackle limited resources in cognitive hierarchies instigated by the IPOMDP. As well as providing healthy strategic detection, perturbing our framework offers insight into mechanisms underlying overly sensitive anomaly detection, evidenced in paranoia, psychosis, and conspiracy theory. We tested the $\aleph$ mechanism in representative mixed-motive and zero-sum Bayesian repeated games. We show that the $\aleph$ mechanism can protect less sophisticated agents against greedy, clever deceivers who utilize nested models to coerce their partners. Such a pretence deployed by a sophisticated sender depends on atypical sequences of actions. Usually, detecting this is outside the scope of a lower DoM agent, but the $\aleph$-mechanism allows a measure of protection. Of course, the higher DoM agent can, at least if well-calibrated with its simpler partner, predict exactly when the $\aleph$-mechanism will fire, and take tailored offensive measures. However, the net effect in both games shows that their hands might be sufficiently tied to make the outcome fairer. Even with the increased deception complexity available to higher DoM agents ($k \geq 2)$ \citep{alon_dis-information_2023}, we expect that agents endowed with the $\aleph$-mechanism, will cope with deceptive ploys in a similar manner to the blueprints presented in this work. We leave this evaluation for future work.

The two defining points for the $\aleph$-IPOMDP framework are augmented inference (the $\aleph$-mechanism), and an appropriate response  (the $\aleph$-policy). For the inference, there could be, as we present here, exquisitely tailored parameters that perform most competently in a given interaction; however, generalisation is hard.  Future work may explore how to incorporate learning into parameter tuning, making the parameter setting an adaptive process rather than a fixed one. Our framework is built on prior work formalising irritation in the context of the multi-round trust task \citep{hula_model_2018}, in which total non-cooperation is a consequence of high irritation. This is a blunt, and deliberately self-destructive instrument to be a credible and thus an effective deterrent \citep{mcnamara2002credible}. One alternative might be a decision that it is worth investing more cognitive effort, such that the victim increases their  DoM \citep{yu_model-based_2021, alaoui_endogenous_2016}, although this could easily become a cognitively expensive arms race \citep{sarkadi_arms_2023}. Instead, we offer a less computationally intensive manner to allow agents with limited resources to mitigate manipulation. One limitation is that our model cannot reason about the intentions and consequent plans of the deceiver, which may be crucial to facilitate opponent learning (as in the work of \citet{yu_model-based_2021}, where agents can learn how to adapt their recursive level via learning). A DoM$(k)$ agent would benefit from such an ability (say via self-play) in repeated interactions, or from the ability to engage in epistemic planning \citep{belle2023epistemic}. However, a savvy opponent, aware of this learning and planning capacity, can still manipulate the learning process to its benefit \citep{foerster_learning_2018}, although such a setting requires substantially more computation due to the recursive nature of the problem. Another issue, keeping with its roots in competitive economics, is our focus on how lower DoM agents might be exploited. One could also imagine the case that the higher DoM agent \emph{exceeds} expectations by sharing more than the lower DoM agent expects. Although this problem may be solved with an editing of the $\aleph$-policy, it could also be a sign that the higher DoM has a social orientation `baked' into its policy to a greater extent than expected. In this case, the lower DoM agent might want to have the capacity to compensate for the over-fair actions and not break down. Naturally, these mechanisms are prone to excess manipulation, and so would need careful monitoring. On the other hand, in continuation of the Machiavellian Theory-of-Mind origin theory by \citet{byrne_machiavellian_1988}, such a beneficiary behaviour may result from an elaborate ruse, designed to fool the victim until the bitter end (like the evil witch in Hansel and Gretel), and hence the $\aleph$-IPOMDP should also account for last minute betrayal by acting defensively. Overreacting may also arise when the mismatch stems not from strategic manipulation but from simple model error \citep{stahl1994experimental, wright_beyond_2010} or a discrepancy between the actual and assumed prior distributions over components of the opponent, i.e., misalignment. This issue gives rise to several related problems. 
First, as illustrated, with the spirit of the no free lunch theorem, the parameters of the $\aleph$-policy need to balance sensitivity and specificity. Hence, the mechanism might either reduce false alarms at the expense of missing true deception or might be overactive and cause the victim to misclassify truly benign behaviour. The latter could then result in paranoid-like behaviour, offering an alternative explanation to paranoia through over-mentalising \citep{alon2023between}. Erroneous anomaly detection has long been theorised as a potential key precursor to paranoid, delusion-like beliefs (\citet{howes2014schizophrenia}), although to date, few theories have been able to formally translate into appropriate contexts observed in the clinic, which invariably involve social and intentional concerns \citep{barnby2023formalising}. This theory moves toward plurality in providing sufficient and necessary computational mechanisms of paranoid-like inference. A second problem is that the model currently assumes k-level reasoning. This means that the $\aleph$-mechanism is activated by agents with lower than $(k-1)$ DoM level. However, a DoM$(k)$ is fully capable of modelling these agents, as they are part of its nested opponent models. Thus, following the ideas suggested by \citep{camerer_cognitive_2004}, future work may extend the opponent set to include all DoM levels up to $k$. Overreaction may arise from an innocent mistake (for example due to high SotfMax temperature) rather than a deceptive move. To mitigate the severe response,  it might be beneficial to incorporate a confidence level to the $\aleph$-mechanism's output, turning it from a binary vector into a probabilistic estimation of the ``Goodness of fit'' of the observed behaviour to the modelled one. Such relaxation is used in Psychology and language (\citet{collins_we_2020}) to allow agents to quantify the reliability of an utterance. Lastly, an issue with inference arises from our assumption that deceptive behaviour can be detected by measuring deviation from typical behaviour. This assumption is challenged in two ways. First, a well-disguised opponent may imitate the behaviour in such a way that is not deviating enough from the assumed typical behaviour or perfectly masquerading in a way that is undetectable by our proposed mechanism. Second, if agents' behaviour becomes more erratic (associated with a lower inverse temperature in the SoftMax policy), the ability to detect and classify both ``expected'' and ``unexpected'' behaviour will be jeopardized \citep{alon2024mal}. In this case, the anomaly detection should take into consideration the increased probability of off-path behaviour and adapt accordingly. We leave room for future work to solve these issues. Opponent verification is relevant to cybersecurity \citep{obaidat2019biometric}, where legitimate users need to be verified and malevolent ones blocked. However, savvy hackers learn to avoid certain anomalies while still exploiting the randomness of human behaviour. To balance effectively between defence and freedom of use, these systems need to probe the user actively to confirm the user's identity. Our model proposes one such solution but lacks the active learning component, which future work may incorporate. Another important ramification of our work is in the application to AI safety and alignment. Recent interest has flourished concerning the emergence of ToM-like ability in LLMs \citep{ullman_detection_2024, sap_neural_2022, kosinski_theory_2023} and simultaneously the risk this brings with aligning LLM goals with human goals. As argued in our paper and hinted at by \citet{van2020editors}, if these models possess high levels of ToM ability, it is not unlikely that they will use it to manipulate human users if it benefits them \citet{Sabour2025HumanDI}. One way of mitigating such a behaviour is via implementation of the $\aleph$-IPOMDP for this problem. Our model proposes blueprints for intention verification to avoid model tampering. If an LLM is maliciously used to affect the user's beliefs, an $\aleph$-like model, applied on the output of the models, may detect a mismatch between the model's output and the expected one for a given context. Coupled with a proper ``punishment'' mechanism (like blocking users from using the model) it may deter perpetrators from engaging in such social engineering endeavours. The $\aleph$-IPOMDP framework represents a significant advance in protecting agents from deception and manipulation in adversarial environments. By augmenting Bayesian inference with the $\aleph$-mechanism we provide a robust method for less sophisticated agents to detect and respond to potentially harmful behaviour from higher-ToM deceivers, a possibility previously out of reach in a typical k-level hierarchy. The framework's effectiveness underscores its potential for broader applications, including cybersecurity and large language models (LLMs). Future work must address model generalisation, cognitive cost management, and the integration of active learning components to further enhance the framework's robustness and adaptability. As we continue to explore and refine these mechanisms, the potential for creating fairer and more secure interactive systems becomes increasingly attainable.

\section{Appendix}

\subsection{Derivation of the Value Function Gradient w.r.t Policy}
\label{app:derivation of value gradient by policy}
\paragraph{Immediate Effect}
We compute the gradient of $\mu$'s value function $V^{\pi^t_{\mu_k}}(b_{\mu_k}(\theta^t_{\nu_{k-1}}))$ in steps. First, we compute the \emph{immediate} effect: changes to $\mu$'s utility function at time $t$ resulting from changes in $\pi^t_{\mu_k}$
\begin{equation}
    \frac{\partial E_{\pi^t_{\mu_k}}[E_{\pi^t_{\nu_{k-1}}}[u^t_\mu(a^{t}_\mu,a^{t}_\nu)]]}{\partial \pi^t_{\mu_k}} = 
    \frac{\partial \sum_{a_\mu^t}\sum_{a^t_\nu}u^t_\mu(a^t_\mu, a^t_\nu)P(a^t_\mu|\pi^t_{\mu_k})P(a^t_\nu|\pi^t_{\nu_{k-1}})}{\partial \pi^t_{\mu_k}}
\end{equation}

Since $\pi^t_{\nu_{k-1}}$ is independent of $\pi^t_{\mu_k}$ - the gradient of the expected immediate utility is:
\begin{equation}
    \sum_{a_\mu^t}\big (\frac{\partial P(a^t_\mu|\pi^t_{\mu_k})}{\partial \pi^t_{\mu_k}}\sum_{a^t_\nu}u^t_\mu(a^t_\mu, a^t_\nu)P(a^t_\nu|\pi^t_{\nu_{k-1}})\big)
    \label{eq:gradient_eq_1}
\end{equation}
Let $\bar{u}(a^t_\mu)$ be the expected utility for $\mu$ from playing action $a^t_\mu$, averaged over $\pi^t_{\nu_{k-1}}$:
\begin{equation}
    \bar{u}(a^t_\mu) = E_{\pi^t_{\nu_{k-1}}}[u^t_\mu(a^{t}_\mu,a^{t}_\nu)]
    \label{eq:gradient_eq_2}
\end{equation}
Plugging Eq. ~\ref{eq:gradient_eq_2} into Eq. ~\ref{eq:gradient_eq_1} yields:
\begin{equation}
    \sum_{a_\mu^t}\frac{\partial P(a^t_\mu|\pi^t_{\mu_k})}{\partial \pi^t_{\mu_k}}\bar{u}(a^t_\mu)
    \label{eq:final_immediate_gradient}
\end{equation}

\paragraph{Gradual Effect}
As changes in $\mu$'s behaviour are propagated through $\nu$'s belief and consequently, $\nu$'s behaviour, we express $\mu$'s belief with its full form (Eq. ~\ref{eq:detailed_marl_recursive_belief_update}) - $b_{\mu_k}(\theta^{t+1}_{\nu_{k-1}}) = p(b_{\nu_{k-1}}^{t}| h^{t-1}) \times p(\psi_\nu\rangle|h^{t-1})$. Since $\psi_\nu$ is independent of $\mu$'s actions, the gradual changes to $\mu$'s value function result from changes to $\nu$'s beliefs.
\begin{align}
    &\gamma \frac{\partial E_{\pi^t_{\mu_k}}[E_{\pi^t_{\nu_{k-1}}}[V^{\pi^{t+1}_{\mu_k}}(b_{\mu_k}(\theta^{t+1}_{\nu_{k-1}}))]]}{\partial \pi^{t}_{\mu_k}} = 
    \gamma E_{\pi^t_{\nu_{k-1}}}[\frac{\partial E_{\pi^t_{\mu_k}}[V^{\pi^{t+1}_{\mu_k}}(b_{\mu_k}(\theta^{t+1}_{\nu_{k-1}}))]}{\partial \pi^{t}_{\mu_k}}] = 
    \\
    \nonumber
    &\gamma\sum_{a^t_\nu}P(a^t_\nu|\pi^t_{\nu_{k-1}})\sum_{a^t_\mu}\frac{\partial P(a^t_\mu|\pi^t_{\mu_k})}{\partial \pi^{t}_{\mu_k}}V^{\pi^{t+1}_{\mu_k}}(b_{\mu_k}(\theta^{t+1}_{\nu_{k-1}})) + 
    P(a^t_\mu|\pi^t_{\mu_k}) \frac{\partial V^{\pi^{t+1}_{\mu_k}}(b_{\mu_k}(\theta^{t+1}_{\nu_{k-1}}))}{\partial \pi^{t}_{\mu_k}}
    \label{eq:value_wrt_to_current_policy_step_1}
\end{align} The first equation stems from the independence of $\nu$'s policy at time $t$ is of $\mu$'s policy at the same time. The 2nd equation is application of the product rule derivation.

We now expand the term $\frac{\partial V^{\pi^{t+1}_{\mu_k}}(b_{\mu_k}(\theta^{t+1}_{\nu_{k-1}}))}{\partial \pi^{t}_{\mu_k}}$. Recall (from Eq. ~\ref{eq:value_function_mu_marl}) that:
\begin{equation}
    V^{\pi^{t+1}_{\mu_k}}(b_{\mu_k}(\theta^{t+1}_{\nu_{k-1}})) = E_{\pi^{t+1}_{\mu_k}}[E_{a^{t+1}_\nu \sim \hat \pi^{t+1}_{\nu_{k-1}}(\psi_\nu, b_{\nu_{k-1}}(\theta^{t+1}_{\mu_{k-2}}))}\big [u_\mu(a_\mu^{t+1}, a_\nu^{t+1}) + \gamma V^{\pi^{t+2}_{\mu_k}}(b_{\mu_k}(\theta^{t+2}_{\nu_{k-1}}))]\big]
\end{equation} Since the actions of $\nu$ at time $t+1$ are a function of $\mu$'s actions at time $t$, we compute the gradient of $\pi_\nu^{t+1}$ w.r.t $\pi^t_\mu$:
\begin{equation}
    \frac{\partial V^{\pi^{t+1}_{\mu_k}}(b_{\mu_k}(\theta^{t+1}_{\nu_{k-1}}))}{\partial \pi^{t}_{\mu_k}} =     
    \frac{\partial V^{\pi^{t+1}_{\mu_k}}(b_{\mu_k}(\theta^{t+1}_{\nu_{k-1}}))}{\partial \pi^{t+1}_{\nu_{k-1}}}
    \frac{\partial \pi^{t+1}_{\nu_{k-1}}}{\partial \pi^{t}_{\mu_k}}    
\end{equation}
Since the policy of $\nu$ is a function of its beliefs, we expand the computation to account for belief updates resulting from $\mu$'s actions:
\begin{equation}
    \frac{\partial V^{\pi^{t+1}_{\mu_k}}(b_{\mu_k}(\theta^{t+1}_{\nu_{k-1}}))}{\partial \pi^{t}_{\mu_k}} =     
    \frac{\partial V^{\pi^{t+1}_{\mu_k}}(b_{\mu_k}(\theta^{t+1}_{\nu_{k-1}}))}{\partial \pi^{t+1}_{\nu_{k-1}}}    
    \frac{\partial \pi^{t+1}_{\nu_{k-1}}}{\partial b_{\nu_{k-1}}(\theta^{t+1}_{\mu_{k-2}})}
    \frac{\partial b_{\nu_{k-1}}(\theta^{t+1}_{\mu_{k-2}})}{\partial \pi^{t}_{\mu_k}} 
\end{equation}
In addition, $\nu$'s future beliefs are a recurrent function of its current beliefs and $\mu$'s actions: $b_{\nu_{k-1}}(\theta^{t+1}_{\mu_{k-2}}) = f(b_{\nu_{k-1}}(\theta^{t}_{\mu_{k-2}}), a^t_\mu)$ (from Eq. ~\ref{eq:marl_recursive_belief_update}) - indicating changes to $\nu$'s beliefs are propagated into its future beliefs, and these are manifested in its actions which governs $\mu$'s utility. Overall, the changes to $\mu$'s policy affect its present and future utilities in both ways.

\subsection{Manipulation and Deception}
\label{app:manipulation_and_deception}

Manipulation is defined as ``the act of changing by artful or unfair means so as to serve one's purpose''. That is, manipulation requires two component: (a) change another agent's behaviour and (b) that the change is beneficial to the manipulator. The first condition implies that the agent that is manipulated has some baseline (expected) behaviour in the absent of the manipulation and that the manipulation \emph{causes} the victim to change its behaviour. This is illustrated in in \citet{jaques_social_2019}, where the manipulator is rewarded for the KLD between the expected (manipulation free) and manipulation induced behaviour of the victim. The second condition mandates that this change benefits the manipulator. This is a rather wide definition that encompasses various settings---from cooperative (where the victim may also gain from the manipulation) to competitive environments. In this work we focus on manipulation in competitive setting. This means that the manipulator gains excess utility from the manipulation while the victim receives lower reward, had it not being manipulated. The above definition is missing a key component to successful manipulation---a manipulable victim. Consider the following example --- a trickster learns that they can insert a coin with a rod to a vending machine and once the machine outputs a can of soda it pulls back the coin, tricking the machine to act in its favour. If the  vending machine had a verification mechanism to detect such trickery \footnote{The authors wish to express their rejection of such mischief.} (for example by comparing the number of coins in the machine to the number of cans omitted)---the ruse would fail. We conclude that this third component, detection avoidance, implies that either the manipulator learns quicker than the victim (or that the victim is not adapting at all), or it has the capacity to avoid detection by exploiting limited computational resources of the victim. In this work we show how deception is a sub-class of manipulation, in which the deceiver manipulates the victim by planting false beliefs in the victim's ``mind''. We show how deception meets all requirements of manipulation---induce changes, personal gain and detection avoidance.

\subsection{$\aleph$-mechanism Activation Analysis in IUG}
\label{app:aleph_mechanism_activation_appendix}

The activation of the $\aleph$-mechanism occurs when one of the component is activated. In this section we provide a detailed analysis of this event, as a function of the mechanism's parameters ($\delta$ and $\omega$) and the DoM$(1)$ threshold. This analysis is complementary to the figures depicted in the main body (Fig. ~\ref{fig:iug_aleph_illustration}). The following figure depicts the probability that the $\aleph$-mechanism will fire (y-axis) as a function of the interaction duration (x-axis) and the aforementioned variables. This figure reinforces the findings from the  false positive analysis---the strong typicality component $Z^1$ is more sensitive than the average reward monitoring $Z^2$ on early trials---when engaging with the low threshold sender. On the other hand, the opposite is true against the high threshold sender. This may be due to the latter's inability (desire) to make high offers, while the former (low threshold) may offer higher partitions (thus respecting the expected average reward) but is prone to repeat the same offer multiple times---triggering the strong typicality component.
\begin{figure}[htbp]
    \centering
    \includegraphics[width=0.67\linewidth]{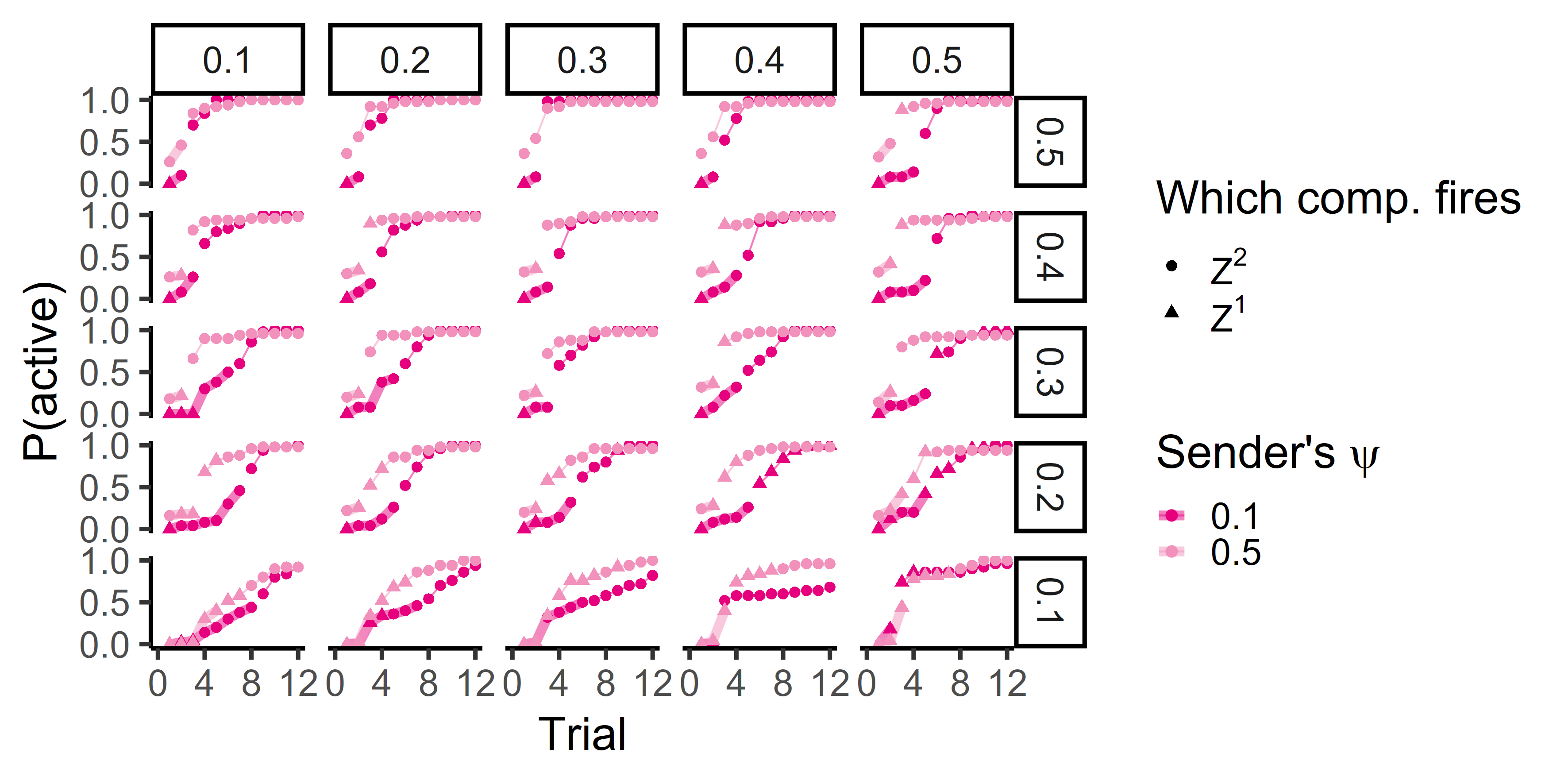}
    \caption{\textbf{Probability of $\aleph$-mechanism activation by component}. The plot shows the probability that the $\aleph$-Mechanism is activated as a function of the trial (x-axis), the sender threshold (colour) and the components (point shape). A change in the line shape (from thick to thin line) indicates a rise in the $\aleph$-mechanism triggering probability above $50\%$. On average, the strong typicality component ($Z^1$) is more likely to trigger the mechanism early on (marked by triangle point) and later activation of the overall mechanism is typically caused by a deviation from the expected reward (monitored by $Z^2$)}
    \label{fig:iug_aleph_mechanism_activation_by_component_threshold_and_trial}
\end{figure}

\subsection{Deception in a Zero-Sum Game}
\label{app:zero_sum_description}
Deception is not limited to mixed-motive games, it may also occur in zero-sum games. A canonical example is Poker \citep{palomaki_machiavelli_2016}, in which players deliberately bluff to lure others into increasing the stakes, only to learn in hindsight that they had been tricked. Simple such games were presented and solved by \citep{zamir_chapter_1992}. Here, we present an RL variant of one of these games \citep[Example 1.3;][]{zamir_chapter_1992}, modelling it with ToM. This illustrates how deceptive behaviour utilises partial information and belief manipulation. Two agents with different DoM levels play the game presented in Fig.~\ref{fig:introduction} (Row/Column game). In this game, one of two payout matrices $G^1, G^2$ (Eq.~\ref{eq:zero_sum_game}) is picked by nature with equal probability, and remains fixed throughout the interaction. The game is played for $T>0$ trials  ($12$ trials in this work). In each trial the row player picks one of two actions $a_\mu \in \{T,B\}$, corresponding to either the top or bottom row, and the column player picks one of the columns: $a_\nu \in \{L,M,R\}$. The agents pick actions simultaneously and observe the action selected by the opponent before the next trial begins. As in the original paper, the payoffs were hidden and revealed only at the end of the game to avoid disclosing the game played. The entries denote the row player payoff which is generated by the  cell that is selected (the column player gets $r_\nu = -r_\mu)$.
\begin{equation}
    G^1 = \begin{pmatrix}
        4 & 0 & 2 \\
        4 & 0 & -2 
    \end{pmatrix},
    G^2 = \begin{pmatrix}
        0 & 4 & -2 \\
        0 & 4 & 2 
    \end{pmatrix}
    \label{eq:zero_sum_game}
\end{equation}

\begin{figure}[htbp]
    \centering
    \includegraphics[scale=.067]{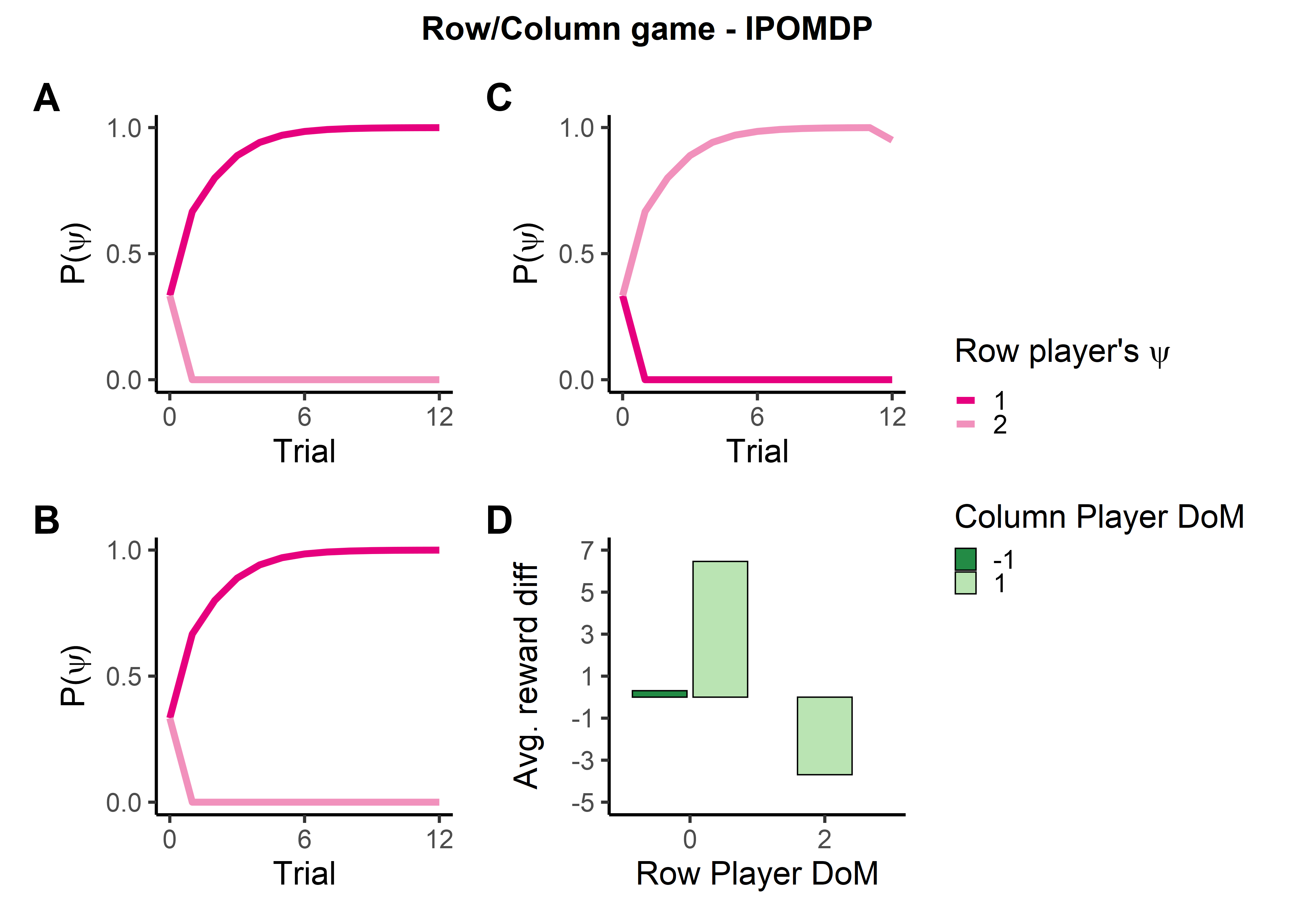}    
    \caption{\textbf{Illustration of deception in Row-Column game}: \textbf{(A)} The DoM$(0)$ column player correctly infers the persona $\psi_{\mu_{-1}}$ of the actions of the DoM$(-1)$ row player. In turn its optimal policy yields it a small, positive reward. \textbf{(B)} Taking advantage of its ToM capacity, the DoM$(1)$ row player manipulates the DoM$(0)$ column player's belief by playing a non typical first move but then changing its policy. The deceptive first move causes the DoM$(0)$ column player to form false beliefs and act to the benefit of the deceptive row player. \textbf{(C)} Reading through the bluff of the DoM$(1)$, the DoM$(2)$ column player correctly identifies the DoM$(1)$ row persona and acts in a counter-deceptive manner. \textbf{(D)} The outcome of these manipulations is presented as the average reward difference between the column and row players' rewards.}
    \label{fig:zero_sum_game_deception_illustration}
\end{figure}

The row player $(\mu)$ may or may not know which matrix is operational (also with equal probabilities), while the column player $(\nu)$ is always ignorant of this fact.
We denote by $\psi_\mu \in \{0,1,2\}$ the row player's persona, with $0$ denoting its ignorance of which matrix was picked, and $1,2$ its respective knowledge. Crucially, the agents receive their cumulative reward only at the end of the game rather than throughout. This implies that $\nu$ can only use $\mu$'s actions to infer this state of the world. Each agent selects its actions using  on a Softmax policy (Eq.~\ref{eq:softmax_distribution}, with a known temperature of $0.1$) based on expected discounted long-term reward. In this game we simulate 2 types of row players: either with DoM level $k \in \{-1,1\}$. We also consider 2 types of $\nu$'s DoM level $\{0,2\}$. The informed DoM$(-1)$ row player ($\psi_{\mu_{-1}} \in \{1,2\}$) assumes the column player is uniform $Q_{\mu_{-1}}(a_\mu,\psi_{\mu_{-1}}) = E_{a_\nu \sim U}[u_\mu(a_\mu,a_\nu)]$. The DoM$(0)$ column player infers a posterior belief about the payoff matrix from the actions of the DoM$(-1)$ row player (Eq.~\ref{eq:dom_zero_belief_update}):
\begin{equation}
    b^t_{\nu_0}(\psi_{\mu_{-1}}) = p(\psi_{\mu_{-1}}|h^{t-1}) \propto P(a_{\mu}^{t-1}|\psi_{\mu_{-1}})p(\psi_{\mu_{-1}}|h^{t-2}) 
    \label{eq:dom_zero_row_player_belief_update}
\end{equation} For example, if the row player constantly plays $T$, this is a strong signal that the payoff matrix is $G^1$, as this action's q-value is $Q_{\mu_{-1}}(T,\psi_{\mu_{-1}}=1) = 2$ compared to $Q_{\mu_{-1}}(B,\psi_{\mu_{-1}}=1) = \sfrac{2}{3}$. This inference is depicted in Fig.~\ref{fig:zero_sum_game_deception_illustration}(A).
Using these beliefs the column player computes the Q-value of each action:
\begin{equation}
    Q^*_{\nu_0}(a^t_\nu,b^t_{\nu_0}(\psi_{\mu_{-1}})) = E_{a^t_\mu \sim \pi_{\mu_{-1}}(\psi_{\mu_{-1}})}[u^t_\nu(a^t_\mu,a^t_\nu)]
    \label{eq:dom_zero_column_player_q_values}
\end{equation}
Its policy then favours the column that yields both it and the row player a $0$ reward ($M$ in $G^1$ and $L$ in $G^2$).

The DoM$(1)$ row player make inferences about the DoM$(0)$ player IRL and plans through its belief update and optimal policy to maximize its reward:
\begin{equation}
    Q^*_{\mu_1}(a^t_\mu,\hat b^t_{\nu_0}(\psi_{\mu_{-1}}), \psi_{\mu_1}) = E_{a^t_\nu \sim \pi_{\nu_{0}}(\theta_{\nu_{0}})}[u^t_\nu(a^t_\mu,a^t_\nu) + \gamma \max_{a^{t+1}_\mu} \{Q^*_{\mu_1}(a^{t+1}_\mu,\hat b^{t+1}_{\nu_0}(\psi_{\mu_{-1}}), \psi_{\mu_1})\}]
    \label{eq:dom_one_row_player_q_values}
\end{equation} notably, these Q-values also take into account the effect of the action on the DoM$(0)$ column player's beliefs. The DoM$(1)$ policy is to trick the DoM$(0)$ column player into  believing a falsehood about the payoff matrix (for example, if the true payoff matrix is $G^1$ it acts in a way typical for a DoM$(-1)$ in $G^2$). This deception utilises the same concepts as in the IUG---the limited opponent modelling of the lower DoM column player and its Bayesian IRL (illustrated in Fig.~\ref{fig:zero_sum_game_deception_illustration}(B)). In turn, the DoM$(0)$ column player's q-value computation (Eq.~\ref{eq:dom_zero_column_player_q_values}) takes as input these false beliefs. This results in its selecting  the column that, instead of yielding it a $0$ reward, is actually the least favourable column ($M$ in $G^1$, $L$ in $G^2$) yielding it a negative utility of $(-4)$. This substantially benefits the deceptive DoM$(1)$ row player, as evident in Fig.~\ref{fig:zero_sum_game_deception_illustration}(D) in terms of the difference in reward between row and column players---in this case $7$ points on average. Lastly, the DoM$(2)$ models the row player as a DoM$(1)$. It inverts its actions to make inference about the payoff matrix from and act optimally, similarly to the DoM$(0)$. Using its nested model of the DoM$(1)$ row player, the DoM$(2)$ column player ``calls the bluff'' and makes correct inferences about the payout matrix Fig.~\ref{fig:zero_sum_game_deception_illustration}(C). Its policy exploits the DoM$(1)$ ruse against itself, by picking the right column, yielding it a reward of $2$ and a reward of $(-2)$ to $\mu_1$. Lacking the capacity to model such counter-deceptive behaviour, the DoM$(1)$ erroneously attributes this behaviour to the SoftMax policy, and its nested beliefs about the column player beliefs are the distorted DoM$(0)$ beliefs. This inability to resist manipulation by the higher DoM column player yields a high income gap, as illustrated in Fig.~\ref{fig:zero_sum_game_deception_illustration}(D).

\begin{figure}[htbp]
    \centering
    \includegraphics[scale=.05]{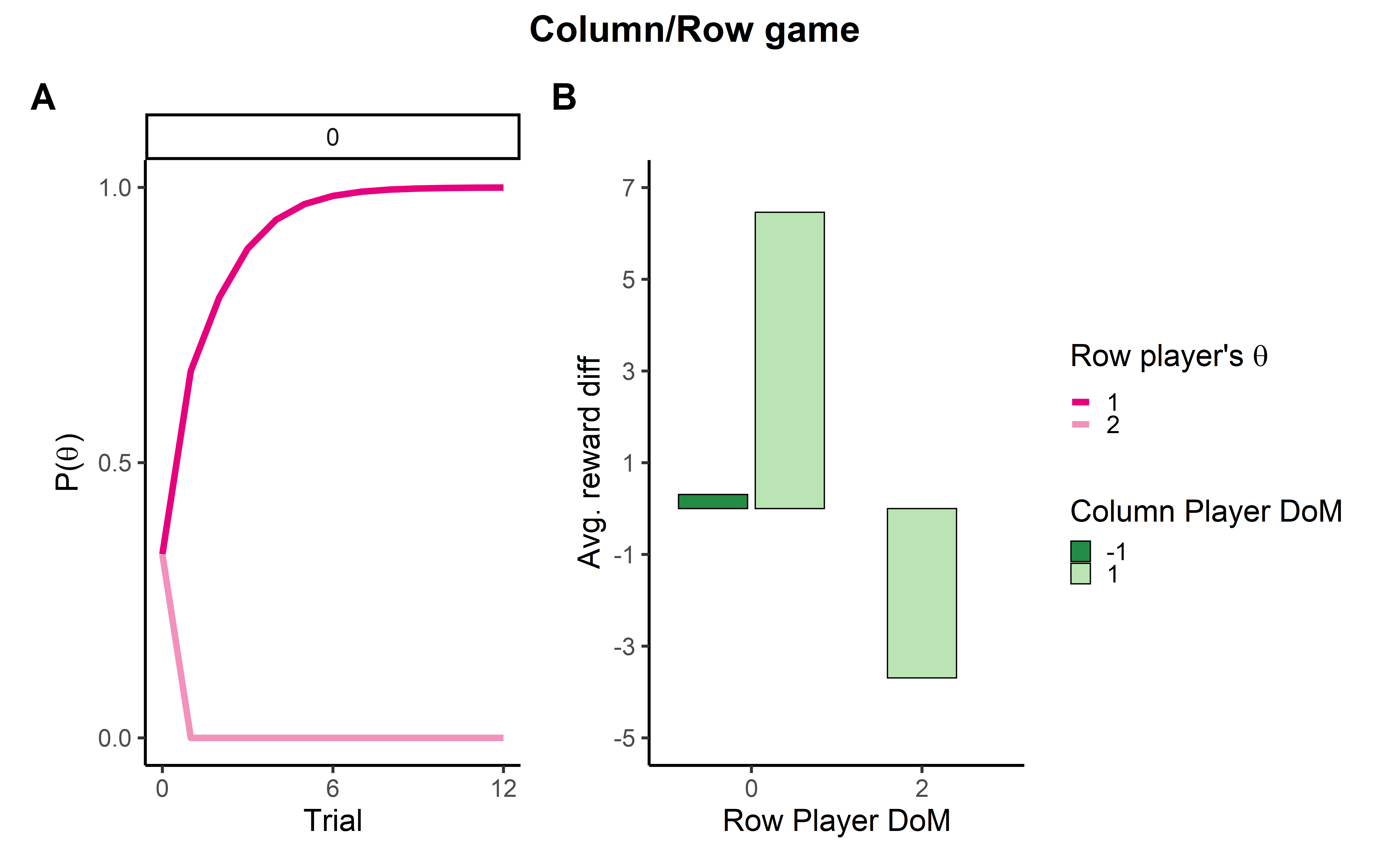}    
    \caption{\textbf{Row/Column zero-sum game}: \textbf{(A)} Illustration of DoM$(0)$ Bayesian-IRL. Given the row player's actions, the DoM$(0)$ column player quickly detects the true type (and payoff matrix). \textbf{(B)} Advantage of high DoM level: in each of the simulated dyads, the higher DoM agent has an edge. Its ability to simulate and predict its opponent's behaviour allows it to gain excess wealth}
    \label{fig:zero_sum_appendix_illustration}
\end{figure} 

\subsubsection{Manipulation, Counter-Manipulation and Counter-Counter Manipulation}
\begin{figure}[htbp]
    \centering
    \includegraphics[scale=.067]{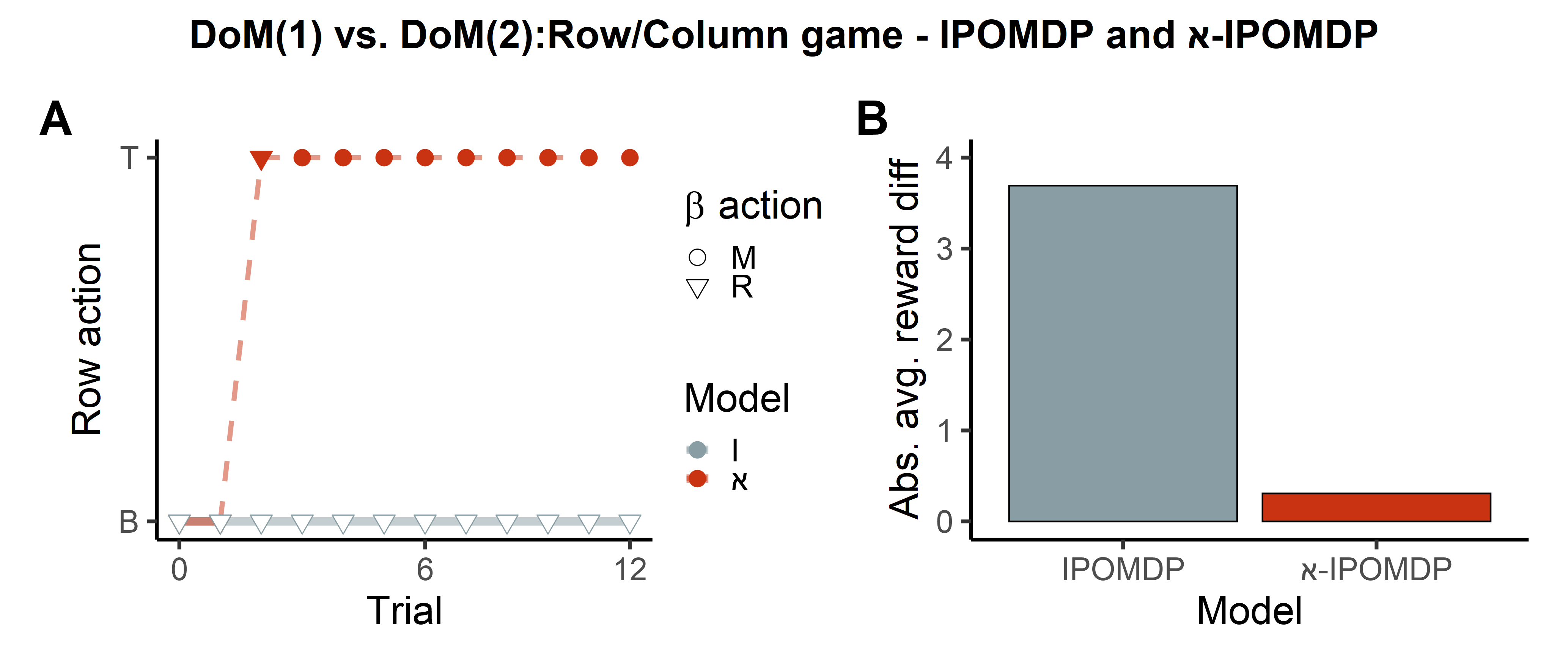}    
    \caption{\textbf{Effect of $\aleph$-IPOMDP in a zero sum game}: \textbf{(A)} Interacting with a $\theta_\mu=1$, the DoM$(0)$ (left column) is deceived by $\mu$'s actions and form false beliefs. However, the DoM$(2)$ utilises its nested model to read through the bluff and correctly identifies $\mu$'s type. \textbf{(B)}(grey line) In turn, the DoM$(2)$ policy exploits of the DoM$(1)$ ruse against it. Augmented with $\aleph$-IPOMDP (red line), the DoM$(1)$ infers this unexpected behaviour as a sign of an external entity, triggering the $\aleph$-mechanism, marked by the dashed line. Its $\aleph$-policy causes the DoM$(2)$ column player to adapt and alter its abusive behaviour. \textbf{(C)} The effect of the $\aleph$-IPOMDP in this task is illustrated via the reduction in the average absolute reward difference.}
    \label{fig:zero_sum_game_main_plot}
\end{figure}

The DoM$(1)$ row player's policy exposes it to potential risk, as it plays the ``risky'' row. For example, in $G^1$ the right column yields it a negative reward. However, given its ability to predict the DoM$(0)$ row player's action this risk is mitigated. As presented above, the DoM$(2)$ column player tricks the trickster by learning to select the right column. The DoM$(2)$ take advantage of the DoM$(1)$ inability to model its behaviour as deceptive and resent it, which yields it a reward of $2$ at each trial, depicted in Fig.~\ref{fig:zero_sum_appendix_illustration}(B). We solve this issue by simulating the game again using the $\aleph$-IPOMDP framework. Due to the reward masking, the DoM$(1)$ detects that they are matched with an external opponent only through the typical-set component. Due to the expected deterministic behaviour of the column player, we use the $\delta$-typicality algorithm in this task. Identifying that they are outmatched, the $\aleph$-policy is to play the MiniMax policy---selecting the row that yields the highest-lowest reward. Interestingly, in this task, this policy is similar to the optimal policy of the ``truth-telling'' DoM$(-1)$ agent. In this case the DoM$(2)$ response is to select the column which yields it the highest reward---namely the one that yields it a zero reward, as evident in \ref{fig:zero_sum_game_main_plot}(C).

There is a similar outcome in the zero-sum game. Here, the behaviour of the deceptive DoM$(2)$ column layer described above would be highly non typical for DoM$(0)$ column player, triggering the $\aleph$-mechanism. The DoM$(1)$ MiniMax $\aleph$-policy, i.e., playing truthfully, causes the DoM$(2)$ to adapt its behaviour appropriately (Fig.~\ref{fig:zero_sum_game_main_plot}(B)). In this case, both parties get $0$ reward, which drops the average absolute reward difference compared to the IPOMDP case, as illustrated in Fig.~\ref{fig:zero_sum_game_main_plot}(C).

\bibliography{additional_refs,references}
\bibliographystyle{unsrtnat}  

\end{document}